%% file: main.tex
\documentclass{article}

\usepackage{PRIMEarxiv}

\usepackage[utf8]{inputenc} 
\usepackage[T1]{fontenc}    
\usepackage{hyperref}       
\usepackage{url}            
\usepackage{booktabs}       
\usepackage{amsfonts}       
\usepackage{nicefrac}       
\usepackage{microtype}      
\usepackage{lipsum}
\usepackage{fancyhdr}       
\usepackage{graphicx}       
\graphicspath{{media/}}     

\usepackage{arydshln}  
\usepackage{mathtools}
\usepackage{amsmath,amssymb,amsfonts}
\usepackage{algorithmic}
\usepackage{graphicx}
\usepackage{textcomp}
\usepackage{xcolor}
\usepackage{ulem}
\usepackage{comment}
\usepackage[ruled,vlined]{algorithm2e}
\usepackage[T1]{fontenc}
\usepackage[utf8]{inputenc}
\usepackage{verbatim}
\usepackage{soul} 
\usepackage[hyphenbreaks]{breakurl}
 \usepackage{float}
\usepackage{hyperref}
\usepackage[hyphenbreaks]{breakurl}
\usepackage{multirow}
\usepackage{adjustbox}

\usepackage{dashrule}

\usepackage{ulem}
\usepackage{enumerate}
\usepackage{colortbl}
\usepackage{booktabs}
\usepackage{array}
\usepackage[english]{babel}
\usepackage{amsthm}

\title{DH-GAN: A Physics-driven Untrained Generative Adversarial Network for 3D Microscopic Imaging using Digital Holography

}

\author{
  Xiwen Chen \thanks{The authors have equal contribution.} , Hao Wang $^*$, Abolfazl Razi \thanks{Corresponding author. email: arazi@clemson.edu} \\
  School of Computing \\
  Clemson University \\
  \And
    Michael Kozicki\\
    School of Electrical, Computer and Energy Engineering \\
    Arizona State University\\
  \And
    Christopher Mann\\
    Department of Applied Physics and Materials Science\\
    Northern Arizona University\\
}

\begin{document}
\maketitle
\begin{abstract}

Digital holography is a 3D imaging technique by emitting a laser beam with a plane wavefront to an object and measuring the intensity of the diffracted waveform, called holograms. 
The object’s 3D shape can be obtained by numerical analysis of the captured holograms and recovering the incurred phase. 
Recently, deep learning (DL) methods have been used for more accurate holographic processing. However, most supervised methods require large datasets to train the model, which is rarely available in most DH applications due to the scarcity of samples or privacy concerns. 
A few one-shot DL-based recovery methods exist with no reliance on large datasets of paired images. Still, most of these methods often neglect the underlying physics law that governs wave propagation. These methods offer a black-box operation, which is not explainable, generalizable, and transferrable to other samples and applications.

In this work, we propose a new DL architecture based on generative adversarial networks that uses a discriminative network for realizing a semantic measure for reconstruction quality while using a generative network as a function approximator to model the inverse of hologram formation. We impose smoothness on the background part of the recovered image using a progressive masking module powered by simulated annealing to enhance the reconstruction quality. The proposed method is one of its kind that exhibits high transferability to similar samples, which facilitates its fast deployment in time-sensitive applications without the need for retraining the network. The results show a considerable improvement to competitor methods in reconstruction quality (about 5 dB PSNR gain) and robustness to noise (about 50\% reduction in PSNR vs noise increase rate).

\end{abstract}


\input{introduction}

\input{related_work}
\input{Methodology}
\input{experiment}

\section{Conclusion}

In this paper, we implemented a GAN-based framework to recover the 3D surface of micro-scaled objects from holographic readings. 
Our method offers several novel features that yield phase retrieval quality far beyond the current practice. 

First, we utilized an AE-based generator network as a function approximator (to map real-valued holograms into complex-valued object waves) in contrast to regular supervised GAN networks, where the generator acts as a density estimator of data samples. Secondly, we implemented a progressive masking method powered by simulated annealing that exploits image foregrounds (e.g., fractal patterns in dendrite samples). This feature facilitates imposing smoothness through TV loss on background areas that further improves the reconstruction and noise removal quality.

The proposed method outperforms both conventional and DL-based methods designed for phase recovery from one-shot imaging under similar conditions. Our method achieves a 10 dB gain in PSNR over the CS-based \cite{zhang2018twin} and about 5 dB gain over the most recent untrained deep learning methods such as DeepDIH \cite{li2020deep}, and DCDO \cite{niknam2021holographic}. An additional 3 dB gain is observed for activating the adaptive masking module. Moreover, our model is sufficiently robust against noise and tolerates AWGN noise up to $\sigma=10$. It shows only about 0.4 dB 
decay per unit noise variance increase, lower than similar methods.
Our method elevates the DL-based digital holography to higher levels with a subtle computation increment. Furthermore, we explored transfer learning to enable fast utilization of the proposed method in time-constrained applications. Our experiments show that using a model trained for a similar sample can offer a reasonable reconstruction quality. Using transfer learning by borrowing network weights trained for a similar sample and performing additional 500 iterations for the new sample brings a considerable gain of about 12 dB compared to independent training with 500 iterations. This observation suggests that the developed model is highly transferrable between samples of the same type, but transferability across different sample types needs further investigation.




\section*{Acknowledgments}
The authors would like to thank Dr. Bruce Gao for his comments on developing the test setup and experiment scenarios. 
This work is supported in part by the USDA AFRI program under grant \#2020-67017-33078.

\bibliographystyle{unsrt}  
\bibliography{references}

\end{document}

%% file: introduction.tex
\section{Introduction}\label{sec:introduction}

\textit{Digital holography} (DH) is a commonly-used technique to exploit the 3D shape of microscopic objects, something not feasible with regular cameras. This powerful technique is used in various applications, including 
micro-particle measurement~\cite{wallace2015robust,patel2021compact}, biology~\cite{xu2001digital}, encryption~\cite{alfalou2009optical}, and visual identification tags \cite{li2020deep}. 
The core idea behind DH is that a laser beam with a plane wavefront experiences diffraction and phase shift when it encounters a microscopic object. 
The interfering wave intensity, also called \textit{hologram}, is captured by a charge-coupled device (CCD) sensor array. The goal of DH is to reconstruct the object's 3D shape by processing the captured holograms \cite{kim2010principles,mann2005high}.

More specifically, the captured hologram $\mathbf{H}(x,y)$ is formed by the superposition of the object wave $\mathbf{O}(x,y)$ and reference wave $\mathbf{R}(x,y)$. This notation assumes that "z" is the propagation direction and x-y is the wavefront plane. In the hologram plane, we have 

\begin{align}\label{eq:1}
    \hat{\mathbf{H}}(x,y) & = |\mathbf{O}(x,y)+\mathbf{R}(x,y)|^2 \\ \nonumber
                    & =(\mathbf{O}(x, y)+\mathbf{R}(x, y))^*(\mathbf{O}(x, y)+\mathbf{R}(x, y))\\
\nonumber
                    & =|\mathbf{O}(x, y)|^{2}+|\mathbf{R}(x, y)|^{2}+\mathbf{R}^{*}(x, y) \mathbf{O}(x, y)+\mathbf{R}(x, y) \mathbf{O}^{*}(x, y),                    
\end{align}
where $(\cdot)^*$ denotes complex conjugation. $\mathbf{O}(x,y)$ and $\mathbf{R}(x,y)$ represent the object wave and reference wave in the hologram plane. 
For the sake of completeness, we can also include a Gaussian zero-mean noise term $\epsilon(x,y)$ to model imaging artifacts, CCD thermal noise, external interference, and other unknown terms, as follows. 
\begin{align}\label{eq:2}
    \mathbf{H}(x,y) = \hat{\mathbf{H}}(x,y)+\epsilon (x,y).
\end{align}

Recovering the object wave from the captured hologram $\mathbf{H}(x,y)$ facilitates 3D-shape reconstruction, due to the linear relationship between the object thickness and incurred phase shift \cite{zeng2021deep}. Therefore, we need to eliminate the zero-order terms $|\mathbf{O}(x, y)|^{2}$ and $|\mathbf{R}(x, y)|^{2}$ and the noise term $\epsilon$ using filtering methods before the phase recovery. This is a fairly easy task in the frequency domain since these zero-order frequency terms are well separated from the modulated terms. The hardest part is extracting $\mathbf{O}(x,y)$ from the two interfering terms ($\mathbf{R}^{*}(x, y) \mathbf{O}(x, y)+\mathbf{R}(x, y) \mathbf{O}^{*}(x, y)$), the so call twin-image problem. Note that $\mathbf{R}^{*}(x, y) \mathbf{O}(x, y)$ and $\mathbf{R}(x, y) \mathbf{O}^{*}(x, y)$ are interchangeably consistent within the solution of Eqs. (\ref{eq:1}) and (\ref{eq:2}) making the inverse problem under-determined.

Generally, there exist two configurations for DH, including \textit{off-axis} and \textit{in-line} (Gabor) digital holography \cite{gabor1948new}. In the former, the incident wave is split into two reference and object waves. These two waves are to be mixed with slightly different arrival angles. This simplifies the subsequent analysis by spatially separating the interfering cross-correlated wave components $\mathbf{O}(x,y)\mathbf{R}^*(x,y)$ and $\mathbf{O}^*(x,y)\mathbf{R}(x,y)$). 
The key drawback of this method is the need for extra equipment to split and mix the reference and object waves as well as accurate calibration of the two laser beams, which hinders making compact DH readers. This method also suffers from potential resolution 
loss by the spatial filtering in the Fourier domain. 

\begin{figure}[h]
\begin{center}
\centerline{\includegraphics[width=1\columnwidth]{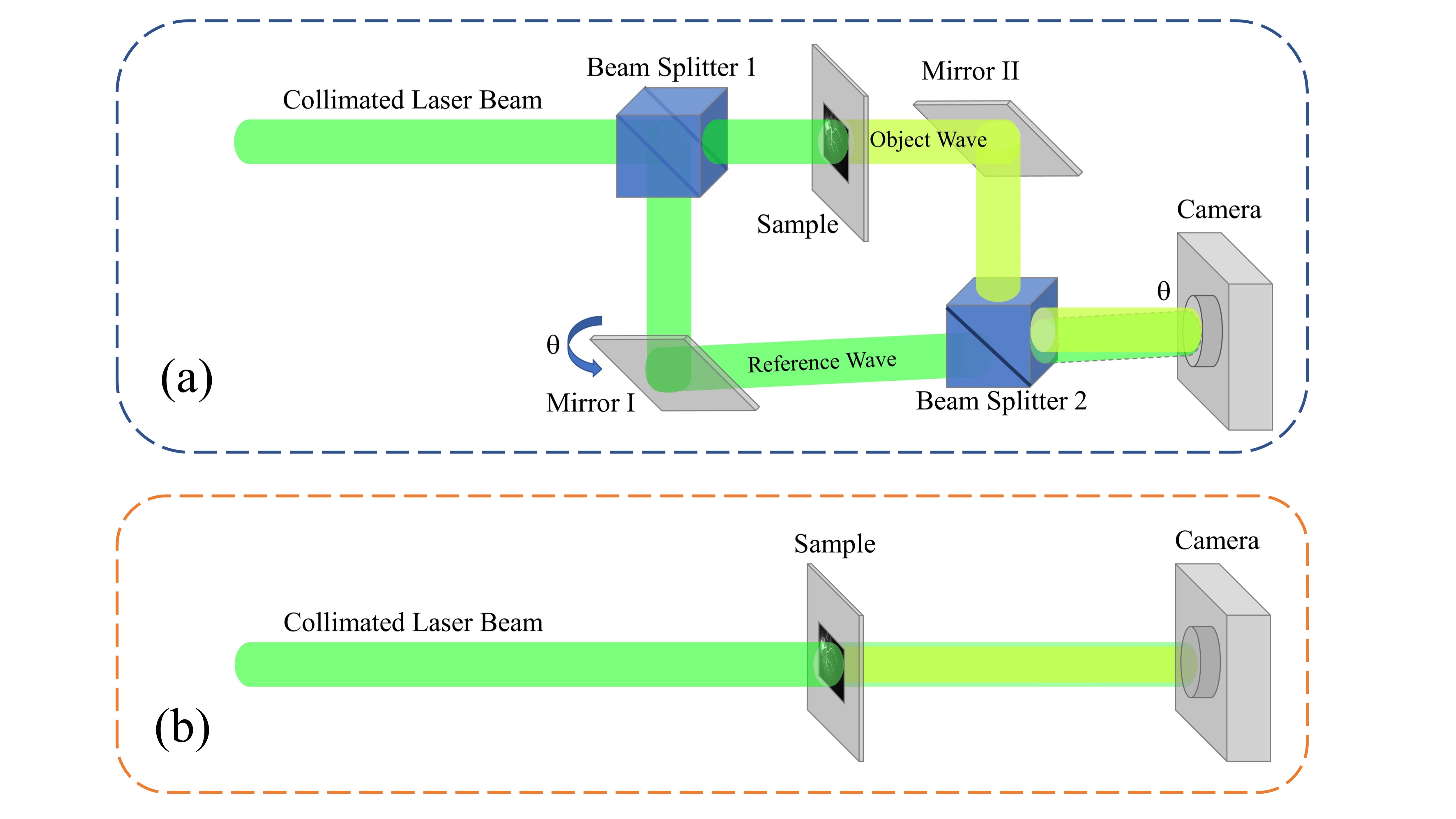}}
\caption{Two typical digital holography setups, including (a) off-axis holography and (b) in-line holography. Setup (b) is used in this project.}
\label{fig:in-line}
\end{center}
\end{figure}

\textit{Digital inline holography} (DIH) entails a much easier hologram rendering method by emitting only one beam through the object and processing the diffracted wave. However, it requires more complex numerical methods for phase recovery to deconvolve the spatially overlapping zero-order and cross-correlated holographic terms. Some methods rely on taking multiple images at different positions to enhance the phase recovery performance \cite{koren1993iterative}. Fig. \ref{fig:in-line} highlights the differences between the two DH imaging configurations.

From a different perspective, image rendering includes reflective and transparent imaging by processing the reflected or passed-through waves, depending on the object's optical characteristics. In this work, we use transparent inline holography (Fig. \ref{fig:in-line}(b)) with single-shot imaging and numerical reconstruction for its more straightforward design and potential for developing low-cost compact and portable readers appropriate for \textit{Internet of Things} (IoT) and supply chain applications \cite{bari2013internet}, especially for \textit{dendritic} tags, our custom-designed visual identifiers \cite{chi2020consistency}.

\subsection{Related Work on Conventional DH Phase Recovery}
The phase recovery problem can be formulated as an inverse nonlinear measurement problem. Several methods have been proposed in the last few years for the numerical phase recovery of holograms. 

The early efforts mainly focused on deterministic and iterative methods based on the wave propagation theory to solve the DH phase-retrieval problem. One of the most successful methods was \textit{transport of intensity} (TIE) used in \cite{teague1983deterministic}.
The basic idea of this method is collecting multiple shots of the hologram at different positions to numerically solve the TIE equation $\frac{\partial \mathbf{H}(x, y)}{\partial z}=-\frac{\lambda}{2 \pi} \nabla_{\perp} \cdot\left(\mathbf{H}(x, y) \nabla_{\perp} \mathbf{\Phi}(x, y)\right)$, where $\mathbf{H}$ denotes the intensity of the hologram, $\nabla_{\perp}$ denotes the gradient operator operating on the dimension $(x,y)$, and $ \mathbf{\Phi}$ denotes the object phase. In addition to the inherent 1st order approximation of the derivative, the TIE-based methods have some limitations. First, the accuracy of the results is conditioned on the precise positioning of the object, and minor deviations may cause unsatisfactory results. Secondly, this method adopts a restrictive presumption that the object substance should be spatially homogeneous \cite{paganin2002simultaneous} 
to use the homogeneous Neumann boundary conditions \cite{zhang2021phasegan}. Iterative solutions for one-shot DIH imaging have no such limitations. Their basic idea is using forward, and backward propagation equations between the imaging and object planes, frequency-domain filtering, and applying physics-based constraints on both hologram space and object space \cite{koren1993iterative,liu1987phase,latychevskaia2007solution}.  


One of the most popular approaches is the Gerchberg-Saxton (GS) algorithm, proposed in \cite{gerchberg1972practical}.
Under this algorithm, at iteration $t$ within the loop, the object distribution $\mathbf{U}_t$ is estimated by back-propagating the current estimated value in the hologram space $|\mathbf{O}_t(x,y)+\mathbf{R}(x,y)|e^{\Phi_{t}}$ to the object plane. Then, the spectral filtering constraints 
are applied to $\mathbf{U}_t$ to refine the object distribution $\mathbf{U}_{t+}$. This new update along with constraints is forward-propagated to the hologram plane to obtain $ |\mathbf{O}_{t+}(x,y)+\mathbf{R}(x,y)|e^{\Phi_{t+}}$. 
This approach is highly reliant on the double-side constraints, and its convergence is not guaranteed, especially when the constraint support regions are not entirely determined \cite{zhang2021phasegan}. 

More recently, a physics-driven \textit{Compressed sensing} (CS) based method has been proposed to solve the twin image problem using single-shot imaging \cite{zhang2018twin}. Specifically, they observed that the real object wave $RR^*O=O$ has sharp edges, while the twin virtual image $RRO^*$ is diffused when mapped to their sparse representation.  
The total variation (TV) loss is applied to the complex-valued object wave to impose sparsity. Moreover, a two-step iterative shrinkage/thresholding (TwIST) algorithm is used to optimize the objective function $ \hat{\mathbf{U}}=\arg \min _{\mathbf{U}}\left\{\frac{1}{2}\|\mathbf{H}-T_f(\mathbf{\mathbf{U}})\|_{2}^{2}+\tau\|\mathbf{U}\|_{t v}\right\}$, where $T_f$ is the forward propagator, $\|.\|_2$ is the 2nd norm, $\|\cdot\|_{t v}$ is the total variation norm, and $\tau$ is a tuning parameter. 
This method is more efficient than the iterative methods, hence is used as a benchmark method in some recent papers \cite{li2020deep,bai2021dual} including our comparative results in this paper. However, it suffers from a few technical issues. For example, imposing explicit sparsity constraints can cause the edge distortion problem. Moreover, the results are sensitive to the choice of $\tau$.

\subsection{Related Work on Deep Learning-based DH}

Recently, deep learning (DL) methods have been used for DH, noting their superior performance in many visual computing and image processing tasks \cite{hassaballah2020deep}. 
In contrast to the conventional phase recovery algorithms that mainly rely on theoretical knowledge and phase propagation models, supervised DL methods often use large-scale datasets for training a black-box model to solve the inverse problem numerically. Therefore, prior knowledge about the propagation model and the system parameters is not necessary to construct DL networks \cite{zeng2021deep}.

For example, the authors of \cite{wang2018eholonet,horisaki2018deep,rivenson2018phase,zhang2020deep} used convolutional neural networks (CNN) with proper regularization terms to reconstruct the object wave from the captured hologram. These method take advantage of the CNN's capability in developing multi-frequency and multi-scale feature maps.  
A generative adversarial network (GAN) is proposed in \cite{wu2019bright} to generate bright-field microscopy at different depths free of the artifacts and noise from the captured hologram. The GAN network learns the statistical distribution of the training samples. Although their one-shot inference was fast, the training time was fairly long, using about 6,000 image pairs (30,000 pairs after data augmentation).

Supervised DL methods, including the aforementioned methods, offer superior performance in phase recovery. Nevertheless, they usually suffer from the obvious drawback of reliance on relatively large datasets for training purposes. For instance, the models in \cite{wang2018eholonet,horisaki2018deep} require about 10,000 training pairs. This requirement becomes problematic since such huge DH datasets rarely exist for different sample types.    
Even if such datasets exist, the training time can be prohibitively long for time-sensitive applications. For instance, the training time with a typical GPU: GeForce GTX 1080 is about 14.5 hours for the model proposed in \cite{rivenson2018phase}. Since the training process is not transferable and should be repeated for different setups and sample types, such a long training phase is not practically desirable. 
In some other applications, such as authenticating objects using nano-scaled 3D visual tags, data sharing can be prohibited for security reasons \cite{chi2020consistency}.

To address the scarcity of paired DH samples, some recent works utilize unpaired data (unmatched holograms and samples) to train their network  \cite{yin2019digital}. 
Specifically, a cycle-generative adversarial network (CycleGAN) 
is employed in \cite{yin2019digital} to reconstruct the object wave from the hologram by training the model with holograms (denoted as domain $\mathcal{X}$) and unmatched objects (denoted as domain $\mathcal{Y}$). Particularly, two generators are used to learn the functions $\mathcal{X}\rightarrow\mathcal{Y}$ and $\mathcal{Y}\rightarrow\mathcal{X}$. A consistency loss is used to enforce the training progress $\mathcal{X}\rightarrow\mathcal{Y}\rightarrow\hat{\mathcal{X}}\approx\mathcal{X}$. 
A similar method based on CycleGAN, called PhaseGAN, is proposed in \cite{zhang2021phasegan}, which used unpaired data for training. The near-field Fresnel propagator~\cite{jenkins1957fundamentals} is employed as part of their framework. Although these methods do not require matched object-hologram samples, they still need large datasets of unmatched hologram samples in the training phase.

Considering the difficulties of developing large DH datasets, some attempts have been made recently to create unsupervised learning frameworks \cite{li2020deep,heckel2018deep,niknam2021holographic}. Most of these frameworks utilize CNN architectures as their backbones since they can capture sufficient low-level image features to reproduce uncorrupted and realistic image parts \cite{ulyanov2018deep}. 
Often, a loss function is employed to minimize the distance between the captured hologram and the artificial hologram obtained by forward-propagating the recovered object wave. For example, our previous work \cite{li2020deep} uses an \textit{hourglass} encoder-decoder structure to reconstruct the object wave from DIH holograms. Inspired by the Deep decoder concept proposed in \cite{heckel2018deep}, the reconstruction algorithm in \cite{niknam2021holographic} 
abandoned the encoder part and only used a decoder with a fixed random tensor as its input. Some classical regularization methods, such as total variation (TV) loss and weight decay, are applied to partially solve the noisy and incomplete signal problem. PhysenNet used a U-net architecture \cite{ronneberger2015u} to retrieve the phase information \cite{wang2020phase}. Most recently, an untrained CNN-based network is employed in dual-wavelength DIH, which benefits from the CNN's capability of image reconstruction and denoising and the Dual-wavelength setup's capability of phase unwrapping \cite{bai2021dual}.

\subsection{Summary of Our Contributions} 

Despite their innovative design and reconstruction efficiency, most of these methods suffer from critical shortcomings. 
First, these untrained networks often use a loss function based on the mean-squared errors (MSE), $L2$-norm, or similar distance measures between the captured hologram and the reproduced hologram. 
This class of loss functions is not capable of measuring structural similarities \cite{palubinskas2017image} and is not fully consistent with the human perception. The perceptual loss, proposed in \cite{johnson2016perceptual}, uses a pre-trained feature extraction backbone to measure the loss, as a reasonable solution for this matter. 
Inspired by this work in developing a semantic similarity measure, we propose an untrained and physics-driven learning framework based on GAN architecture for one-shot DH reconstruction. In our method, the \textit{discriminator network} contributes a learnable penalty term to evaluate the similarity between the reproduced and the captured holograms. As we will discuss later in section \ref{sec:loss}, the role of the generator network in our network is textit{function approximator} to model the inverse of the hologram generation process (i.e., mapping the hologram to complex-valued object wave), as opposed to general GANs, where the generator network learns the data distribution to create new samples from noise.

Another drawback of most aforementioned DL-based methods is their lack of interpretability and ignorance of physics knowledge. Therefore, there are always two risks (i) over-fitting and (ii) severe performance degradation under minor changes to sample characteristics and test conditions.  
We address these issues in two different ways. First, we incorporate forward and backward propagation into our model, following some recent works \cite{li2020deep,niknam2021holographic,wang2020phase}. Secondly, we implement a new spatial attention module using an adaptive masking process to split the object pattern into foreground and background regions and impose smoothness on the image background. The background mask update is performed based on the reconstructed object wave quality to be regulated by \textit{simulated annealing} (SA) optimization to start from more aggressive updates and settle with more conservative changes when the network is converged.     
Imposing smoothness constraint on the gradually-evolving background area, makes our method fundamentally different than some iterative methods that enforce physics-driven hologram formation equations on the support region (i.e., the foreground) \cite{gerchberg1972practical,zalevsky1996gerchberg} or the entire image \cite{latychevskaia2019iterative}. 


We show that our framework is generic and independent of the choice of the generator network. In particular, we tested our framework with two recently developed generators, the fine-tuned version of DeepDIH \cite{li2020deep} and the deep compressed object decoder (DCDO) \cite{niknam2021holographic}. We also show that adding a super-resolution layer to the utilized auto-encoder (AE) improves the quality of the phase recovery.

This paper is organized as follows. Section II reviews the hologram formation process and recasts it as a nonlinear inverse problem. Section III elaborates on the details of the proposed DL method for phase recovery, highlighting its key features and differences from similar methods. Experimental results for simulated holograms, publicly available samples, and our dendrite samples are presented in Section IV followed by concluding remarks in Section V.

\section{Problem Formulation}

The goal of this work is to design an unsupervised physics-driven DL network to reconstruct the 3D surface of microscopic objects, especially \textit{dendrites}, micro-scaled security tags used to protect supply chains against cloning and counterfeit attacks (see Section \ref{sec:dendrite} for details of \textit{dendrites}).

The incident wave passing through a thin transparent object can be characterized as a complex-valued value
\begin{align} \label{eq:perturb}
\mathbf{O}(x,y;z=0)= \mathbf{R}(x,y;z=0)t(x,y),
\end{align}
where $\mathbf{R}(x,y;z=0)$ is the reference wave (i.e., the incident wave if the object is not present) and $t(x,y)=A(x,y)\text{exp}(\phi(x,y))$ is the incurred perturbation term caused by the object. $t(x,y)$ includes attenuation $A(x,y)$ and phase shift $\phi(x,y)$ \cite{latychevskaia2015practical}. After the Fresnel diffraction at distance $z=d$, the forward-propagated object wave $\mathbf{O}(x,y;z=d)$ is formed as
\begin{align} \label{eq:prop1}
    \mathbf{O}(x,y;z=d) &= \mathbf{p}(\lambda,z=d)\circledast \mathbf{O}(x,y;z=0) \\ \nonumber
                             & = \mathcal{F}^{-1}\{\mathbf{P}(\lambda,z=d)\cdot\mathcal{F}\{\mathbf{O}(x,y;z=0)\}\},
\end{align}
where $\lambda$ represents the wavelength and $\circledast$ is the convolution operator. $\mathcal{F}\{\cdot\}$ and $\mathcal{F}^{-1}\{\cdot\}$ denote the direct and inverse Fourier transforms, respectively. Here, $\mathbf{P}(\lambda,z)= \mathcal{F}\{\mathbf{p}(x,y,z)\}$ is the transmission function, defined as
\begin{align}  \label{eq:prop2}
  \mathbf{P}(\lambda,z) = \exp \left(\frac{2 \pi j z}{\lambda} \sqrt{1-\left(\lambda f_{x}\right)^{2}-\left(\lambda f_{y}\right)^{2}}\right), 
\end{align}
where $f_{x}$ and $f_{y}$ denote the spatial frequencies. The formed hologram in the detector plane is

\begin{align}   \label{eq:hologram}
    \mathbf{H}(x,y;\lambda,z) &= |\mathbf{p}(\lambda,z=d)\circledast (\mathbf{O}(x,y;z=0)+\mathbf{R}(x,y;z=0))|^2.
\end{align}
Our ultimate goal is to recover the object-related perturbation $t(x,y)$ or equivalently the complex-values object wave $\mathbf{O}(x,y)$ from the captured hologram $\mathbf{H}(x,y)$, that is consistent with Eqs. (\ref{eq:perturb}-\ref{eq:hologram}).



\begin{figure}[htbp]
\begin{center}
\centerline{\includegraphics[width=0.8\columnwidth]{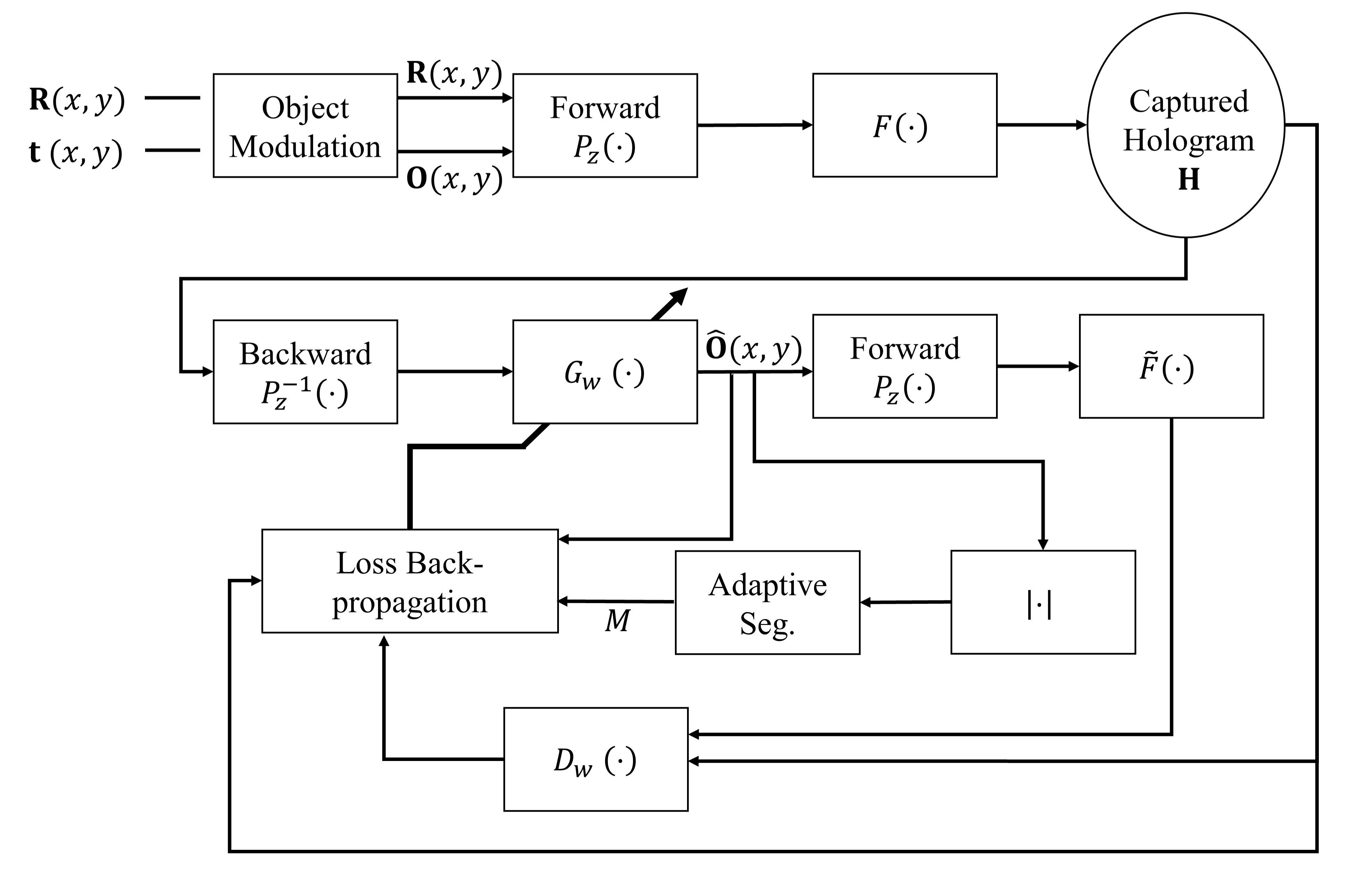}}
\caption{The overall block diagram of the hologram formation along with the proposed DL architecture for phase recovery.} 
\label{fig:block}
\end{center}
\end{figure}

%% file: Methodology.tex
\section{Proposed Method}\label{sec:method}


The essence of our method relies on using a GAN-based architecture with several key modifications. More specifically, consider a chain $\mathbf{O} \stackrel{F(\cdot)}{\longrightarrow} \mathbf{H_0} \stackrel{P_z(\cdot)}{\longrightarrow} \mathbf{H} \stackrel{P_z^{-1}(\cdot)}{\longrightarrow} \mathbf{H_0 }
\stackrel {G_W(\cdot)}{\longrightarrow}\mathbf{ \hat{O}} \stackrel{\Tilde{F}(\cdot)}{\longrightarrow} \mathbf{\hat{H}_0} \stackrel{P_z(\cdot)}{\longrightarrow} \mathbf{\hat{H}}$ (Fig. \ref{fig:block}), 
where $\mathbf{O} \in \mathbb{R}^{  h \times  w \times 2} $ is the inaccessible and unknown complex-valued object wave with height $h$ and width $w$, $\mathbf{H_0} \in \mathbb{R}^{h \times  w \times 1}$ is the produced hologram in the object plane, and $\mathbf{H} \in \mathbb{R}^{h \times  w \times 1}$ is the hologram in the sensor plane. Similarly, $\mathbf{\hat{O}}$, $\mathbf{\hat{H}}_0$, $\mathbf{\hat{H}}$, are the reconstructed versions of the object wave, the hologram in the object plane, and the hologram in the sensor plane. Forward and backward angular spectrum propagation (ASP) according to Eqs. (\ref{eq:prop1}) and (\ref{eq:prop2}) are represented by $P_z(\cdot)$ and $P_z^{-1}(\cdot)$. Likewise, $F(\cdot): \mathbb{R}^{h \times  w \times 2} \mapsto \mathbb{R}^{h \times  w \times 1}$ represents the hologram formation according to Eqs. (\ref{eq:perturb})-(\ref{eq:hologram}). Our goal is to develop a generator network $G_w(\cdot): \mathbb{R}^{h \times  w \times 1} \mapsto \mathbb{R}^{h \times  w \times 2}$ that models the inverse of the hologram formation process to reproduce the object wave $\mathbf{\hat{O}}$ as close as possible to $\mathbf{O}$ under some distance measure $d(\mathbf{\hat{O}},\mathbf{O})$. However, we can not quantify $d(\mathbf{\hat{O}},\mathbf{O})$ since $\mathbf{O}$ is inaccessible. To address this issue and noting that the hologram formation process $F(\cdot)$ is known, we apply the same process to the reconstructed wave $\mathbf{\hat{O}}()$ to obtain a corresponding reproduced hologram $\mathbf{\hat{H}}$. Then, we use the surrogate distance $d(\mathbf{\hat{H}},\mathbf{H})$ instead of $d(\mathbf{\hat{O}},\mathbf{O})$ to assess the reconstruction quality. 
Finally, note that we used $\Tilde{F}(\cdot)$ for numerical hologram formation to account for minor differences with the real hologram formation for parameter mismatch $\lambda, z$, and for adopting some idealistic assumptions (e.g., plane wavefront).

\begin{figure}
\begin{center}
\centerline{\includegraphics[width=1\columnwidth]{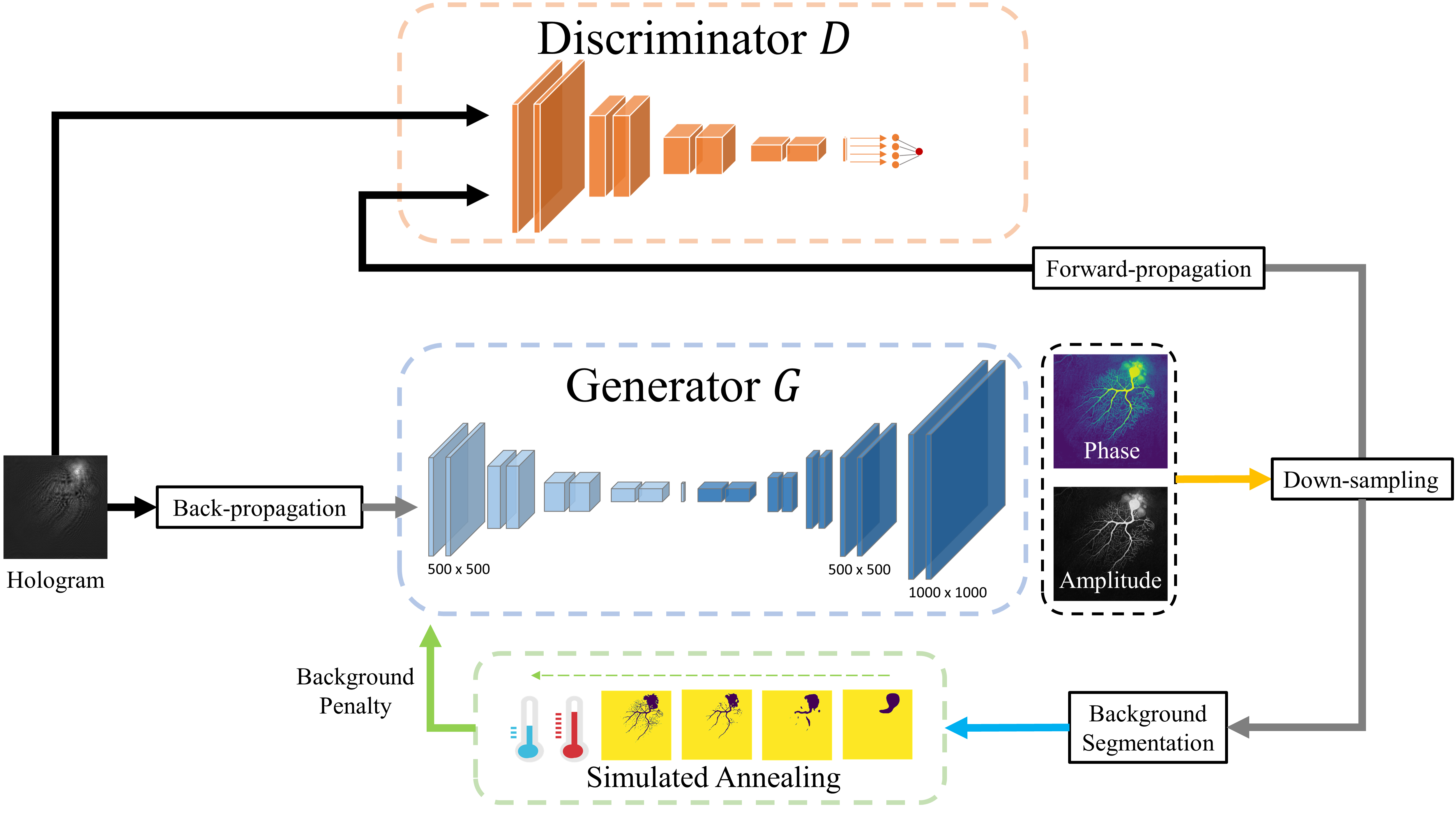}}
\caption{The overall framework of our untrained GAN-based network which consists of AE-based generator network $G$, a discriminator network $D$, and a SA-based adaptive masking module.}
\label{fig:G_SR}
\end{center}
\end{figure}

\subsection{Optimization through Loss Function}\label{sec:loss}

Figs. \ref{fig:block} and \ref{fig:G_SR} present the details of the proposed DL Architecture for DIH phase recovery. 
The loss term used in the generator network $G_W(\cdot)$ includes the following components.

\begin{itemize}
    
\item One term is the MSE distance between the reproduced and the captured hologram $d_1(\mathbf{\hat{H}},\mathbf{H})=MSE(\mathbf{\hat{H}},\mathbf{H})$ used to directly train the AE-based generator, following the physics-driven methods \cite{li2020deep,niknam2021holographic,wang2020phase}. 
    
\item Noting the limitations of MSE and 2nd norm, we also use a discriminator network $D_W(): \mathbb{R}^{h \times  w \times 2} \mapsto \mathbb{R}^{1}$ to produce a learnable penalty term by maximizing the confusion between the reproduced and captured holograms that can be an indicator of the quality of the hologram. Suppose $D_W(\mathbf{H})$ and $D_W({P_z}(\tilde{F}(G_W(P_Z^{-1}(\mathbf{H}))))$ be the probability of the captured and reproduced holograms being real. Then, we must maximize the first term and minimize the second term when training the discriminator to distinguish between the real and numerically-regenerated holograms. However, we maximize the second term when training the generator to make the reproduced holograms as close as possible to the captured hologram to fool the discriminator. This is equivalent to the conventional GAN formulation 
\begin{equation} \label{eq:GAN}
 \mathcal{L} =\min_{G_W}\max_{D_W}\mathbb{E}_{x\sim P_{data}}\log[D_W(x)]+\mathbb{E}_{z\sim p_z}\log[1-D_W(G_W(z))],
\end{equation}  
with a few modifications. First, the artificially generated output from noise $G_W(z)$ is replaced by ${P_z}(\tilde{F}(G_W(P_Z^{-1}(\mathbf{H}))))$, i.e. the reproduced hologram from the recovered object wave $G_W(\mathbf{H})$. Note that we have $H={P_z}(\tilde{F}(G_W(P_Z^{-1}(\mathbf{H}))))$ for the optimal case of $G_W=\tilde{F}^{-1}=F^{-1}$. 
Secondly, the role of the generator in our design is not modeling the output distribution to produce seemingly real outputs from noise-like inputs; instead, it is used as a function approximator to map the backward-propagated hologram to object wave $\mathbf{H}_0 \stackrel{G_W(\cdot)}{\longrightarrow} \mathbf{\hat{O}}$. Therefore, we can drop the expectation operation $\mathbb{E}$ from the optimization problem in Eq. (\ref{eq:GAN}). 

\item Finally, to incorporate our prior knowledge, we use a new term that imposes smoothness on the image background. This is to embrace the fact that in most real scenarios, the samples are supported with a transparent glass slide, meaning that the background of the reconstructed object should present no phase shift. In other words, $t(x,y)=A(x,y)e^{\Phi(x,y)}=1 \Rightarrow \mathbf{O}(x,y)=\mathbf{R}(x,y)$ based on Eq. (\ref{eq:perturb}), which means zero phase shift in the object wave for all pixels out of the object boundary, $(x,y)\notin \mathcal{S}$. This approach is inspired by the physics-informed neural networks (PINN) \cite{cai2022physics} that use boundary conditions to solve partial differential equations (PDEs). 
Our approach to detecting image background is discussed in Section \ref{sec:mask}.    

\end{itemize}

To summarize, the proposed network aims to solve the following optimization problem: 
\begin{align}\label{eq:loss}
    \mathcal{L}=\min _{G_W} \max _{D_W} \log [D_W(\mathbf{H})]+ \log [1-D_W({P_z}(\tilde{F}(G_W(P_Z^{-1}(\mathbf{H})))))]+\lambda_1 \mathcal{L}_{Auto}(G_W(\mathbf{H}))+\lambda_2 \mathcal{L_{B}}(G_W(\mathbf{H})).
\end{align}

The first two terms represent the GAN framework loss with the ultimate goal of making the generator $G_W()$ as close as possible to the inverse of hologram formation $F^{-1}()$ through iterative training of the generator and discriminator networks. We have used an auto-encoder architecture for the generator following our previous work \cite{li2020deep}, whose loss function is represented by $\mathcal{L}_{Auto}$. Likewise, $\mathcal{L_{B}}$ represents the background loss term for points out of the object mask $p \notin \mathcal{S}$ with $\lambda_1$ and $\lambda_2$ being tuning parameters.


In the training phase, the loss of $G_w$ and $D_w$ are minimized sequentially,
\begin{align}
    \mathcal{L}_{G_w} &= \min_G  \log [1-D_W({P_z}(\tilde{F}(G_W(P_Z^{-1}(\mathbf{H})))))]+\lambda_1 \mathcal{L}_{Auto}(G_W(\mathbf{H}))+\lambda_2 \mathcal{L_{B}}(G_W(\mathbf{H})) \\  \nonumber
    \mathcal{L}_{D_w} &= \max_D\log [D_W(\mathbf{H})]+ \log [1-D_W({P_z}(\tilde{F}(G_W(P_Z^{-1}(\mathbf{H})))))].
\end{align}

To avoid the lazy training of the generator and achieve larger gradient variations, especially at the beginning training steps, we solve the following equivalent optimization problem
\begin{align}
    \mathcal{L}_{G_w} &= \min_G  -\log [D_W({P_z}(\tilde{F}(G_W(P_Z^{-1}(\mathbf{H})))))]+\lambda_1 \mathcal{L}_{Auto}(G_W(\mathbf{H}))+\lambda_2 \mathcal{L_{B}}(G_W(\mathbf{H})).
\end{align}

Since this network has only one fixed input and target, the GAN structure aims to map the input to a reproduced domain as close as possible to the target, even without the $\mathcal{L}_{Auto}$ and $\mathcal{L}_{B}$ terms. Adding these terms enhances the reconstruction quality by enforcing our prior knowledge. Besides, since the discriminator $D_w$ would extract deep features via its multiple convolutional layers, compared with the MSE loss or $L2$ loss, its similarity evaluation would intuitively be more meaningful. Thus, the network would learn a more robust translation from the digital hologram to the object wave.

The auto-encoder loss term $\mathcal{L}_{Auto}(G_W(H))$ in Eq. (\ref{eq:loss}) is used to directly minimize the gap between the captured hologram and the numerically reconstructed hologram, independent from the utilized discriminator. 

\begin{align}\label{eq:auto}
\nonumber
&\mathcal{L}_{Auto}(G_W(\mathbf{H}))=d_{MSE}(\mathbf{H},\mathbf{\hat{H}})=\frac{1}{h \times  w }\|\mathbf{H}-\mathbf{\hat{H}}\|_{2}^{2}, \\
&\mathbf{\hat{H}}=P_z(\tilde{F}(G_W(P_z^{-1}(\mathbf{H})))),
\end{align}
where the captured and reconstructed holograms ($\mathbf{H}$, $\mathbf{\hat{H}}$) are representatives of the AE input and output after proper propagation.

Finally, we use total variation (TV) loss to enforce smoothness on the image background, or simply the pixels $p=(x,y) \notin \mathcal{S}$ out of the region of interest (ROI), or the image foreground. This incorporates our prior knowledge about zero-shift for background pixels beyond ROI, and improves the reconstruction quality. 
The TV loss for complex-valued 2D signal $z$ 
is
\begin{align} \label{eq:LB1}
    \mathcal{L}_{B}(z)&= \int_{z \in \Omega_B} \big(|\nabla\Re(z)|+|\nabla\Im(z)|\big) \mathbf{d}x\mathbf{d}y,
\end{align}
where $\Omega_B$ denotes the support set of $z$, and $\Re(z)$ and $\Im(z)$ denote the real and imaginary parts of $z$, respectively. In our case, the points z are taken from $\tilde{F}(G_W(P_z^{-1}(H)))$ and $\Omega_B=
\{(x,y)|1\leq x \leq w, 1\leq y \leq h, (x,y) \notin \mathcal{S}\}$.

For discrete signals, we use the approximation $|\nabla_x\Re(z)|=|\Re(z)_{x+1, y}-\Re(z)_{x, y}|$. Noting $|\nabla\Re(z)|=(|\nabla_x\Re(z)|_2^2+|\nabla\Re_y(z)|_2^2\big)^{1/2}$, Eq. (\ref{eq:LB1}) converts to
\begin{align} \label{eq:LB2}\nonumber
     \mathcal{L}_{B}(z) =\frac{1}{|\Omega_B|}\sum_{x,y\in \Omega_B} &\big(\left|\Re(z)_{x+1, y}-\Re(z)_{x, y}\right|^2+\left|\Re(z)_{x, y+1}-\Re(z)_{x, y}\right|^2\big)^{1/2} \\
     &+ \big(\left|\Im(z)_{x+1, y}-\Im(z)_{x, y}\right|^2+\left|\Im(z)_{x, y+1}-\Im(z)_{x, y}\right|^2\big)^{1/2}, 
\end{align}
where $|\Omega_B|$ is the cardinality (the number of points) of set $\Omega_B$. For simplicity, we skip the square root operation, and use the following version, which is computationally faster.

\begin{align} \label{eq:LB3}
    \nonumber
     \mathcal{L}_{B} =\frac{1}{|\Omega_B|}\sum_{x,y\in \Omega_B} &\left|\Re(z)_{x+1, y}-\Re(z)_{x, y}\right|^2+\left|\Re(z)_{x, y+1}-\Re(z)_{x, y}\right|^2 \\
     &+ \left|\Im(z)_{x+1, y}-\Im(z)_{x, y}\right|^2+\left|\Im(z)_{x, y+1}-\Im(z)_{x, y}\right|^2. 
\end{align}
The details of the adaptive masking to define the ROI is discussed below.

\subsection{Adaptive Masking by K-means and Simulated Annealing}\label{sec:mask}

\begin{figure}[htbp]
\begin{center}
\centerline{\includegraphics[width=1\columnwidth]{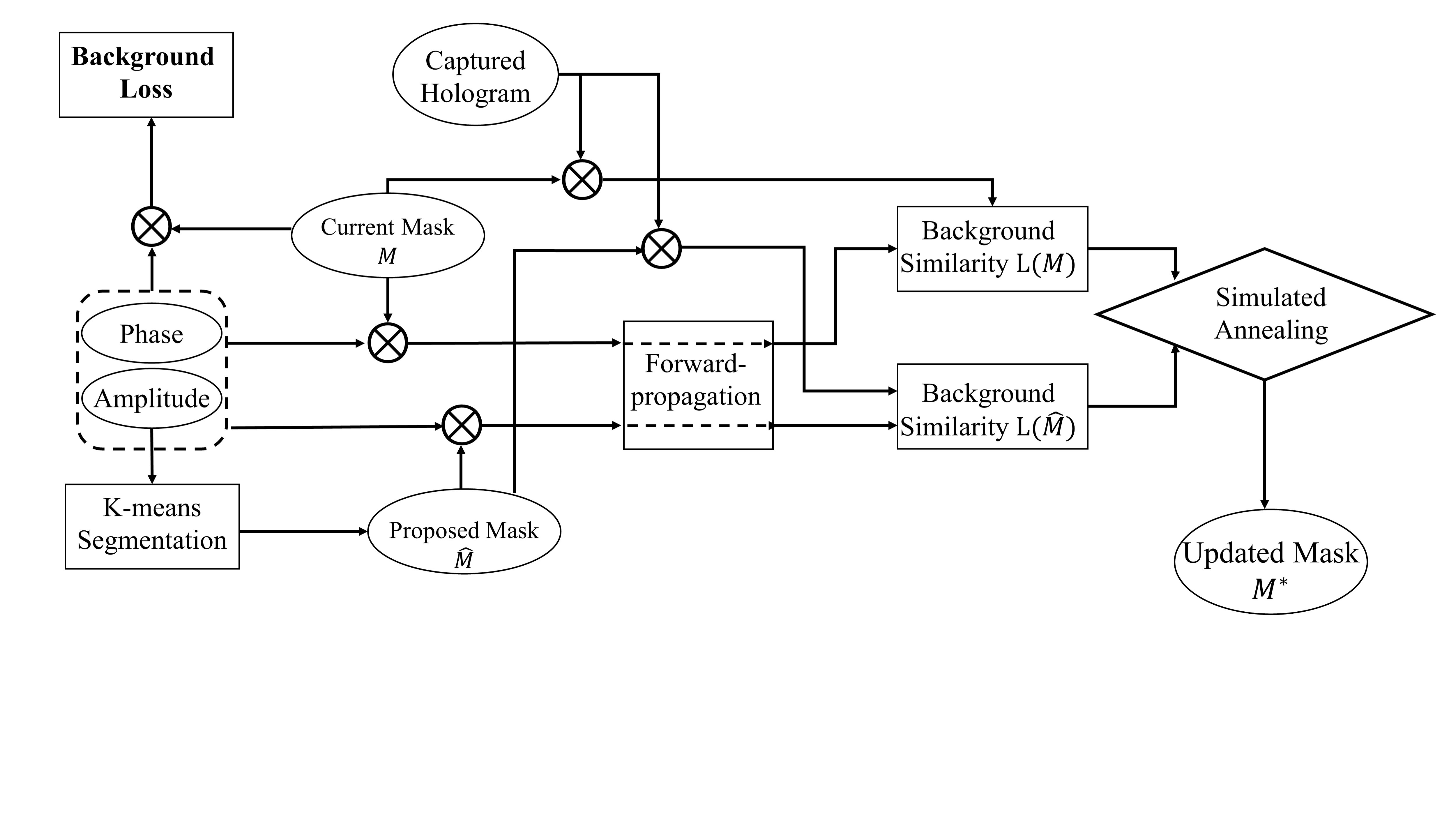}}
\caption{The block-diagram of the adaptive segmentation to create background loss. The operator $\otimes$ denotes element-wise multiplication, indicating that all operations are only applied on the background area. The mask update process is explained in Section \ref{sec:mask}.}

\label{fig:bg_loss}
\end{center}
\end{figure}

The background loss $\mathcal{L}_{B}$ in Eq. (\ref{eq:loss}) operates on the background area of the output image, as shown in Fig. \ref{fig:bg_loss}. The background area is determined by a binary mask $M^{(t)}$, where $t$ is a discrete number denoting the mask update time point. 
To this end, a binary mask $\hat{M}^{(t)}$ is developed by applying K-means segmentation (with $K$=2) to $|\mathbf{\hat{O}}_0|$, the amplitude of the reconstructed object wave at $z=0$. We consider the resulting mask as a "proposal mask", which may or may not be accepted. Rejection means that we use the previously formed mask $M^{(t-1)}$ to calculate the background loss. To avoid instability of the results and unnecessary mask updates, we use a mechanism that tends to make more frequent (aggressive) updates at the beginning and less frequent (conservative) updates when the algorithm converges to reasonably good results. A natural way of implementing such a mechanism is using \textit{simulated annealing} (SA) algorithm where the variation rate decline is controlled by \textit{temperature} cooling. 

The SA algorithm is initialized by temperature $T_0$ for time $t=0$. We also set the first mask $M^{(0)}=[1]_{h \times w}$, assuming no foreground is detected yet. 
To update the mask at time $t=1,2,3,\dots$, we compare the MSE distance between the reproduced hologram $\hat{H}$ and the captured hologram $H$ on the background areas determined once by the previous mask $M^{(t-1)}$ and next by the current mask proposal $\hat{M}^{(t)}$. Mathematically, we compute $\delta_{t-1} = d_\text{MSE}(\mathbf{H},\mathbf{\hat{H}}; M^{(t-1)})$, and $\hat{\delta}_t = d_\text{MSE}(\mathbf{H},\mathbf{\hat{H}}; \hat{M}^{(t)})$. Inequality $\hat{\delta}_t< \delta_{t-1}$ means that the consistency between the captured and reconstructed holograms improves by using the current mask proposal, so we accept the proposal and update the mask $M^{(t)}=\hat{M}^{(t)}$. Otherwise, we lower the \textit{temperature} as $T_t=T_{t-1}/\log(1+t)$, and then update the mask with Probability $e^{-(\hat{\delta}_t- \delta_{t-1}) / T_t}$. It means that as the time passes, the update probability declines. The summary of this algorithm is presented below.

\begin{algorithm}\label{alg:1}
\caption{Adaptive Background Masking} 
\label{alg1}
\begin{algorithmic}
\REQUIRE $T_0,\mathbf{H}_1,\mathbf{H}_2,\dots,\mathbf{\hat{H}}_1,\mathbf{\hat{H}}_2,\dots,\mathbf{\hat{O}}_1,\mathbf{\hat{O}}_2, \dots$ 
\ENSURE $\hat{M}^{(1)},\hat{M}^{(2)},\hat{M}^{(3)},\dots$
\STATE \noindent \textbf{Initialize:} $M^{(0)}=[1]_{h \times w}$ 
\STATE $t=1$
\WHILE{not converged}
\item $\hat{M}^{(t)}\Leftarrow\mathbf{KMeans}(\left| \hat{O}_t \right|)$
\item $\delta_{t-1} \Leftarrow d_\text{MSE}(\mathbf{H},\mathbf{\hat{H}}; M^{(t-1)})$
\item $\hat{\delta}_t \Leftarrow d_\text{MSE}(\mathbf{H},\mathbf{\hat{H}};\hat{M}^{(t)})$
\item\IF{$\hat{\delta}_t< \delta_{t-1}$} 
\item$M^{(t)}\Leftarrow \hat{M}^{(t)}$
\item \ELSIF{$p\sim \mathcal{U}(0,1)\leq \exp(-(\hat{\delta}_t< \delta_{t-1}) /T_t ) $} 
\item $M^{(t)}\Leftarrow \hat{M}^{(t)}$
\item \ELSE 
\item $M^{(t)}\Leftarrow {M}^{(t-1)}$
\item \ENDIF
\item $T_t \Leftarrow T_{t-1} / log(1 + t)$
\item $t\Leftarrow t+1$
\ENDWHILE
\end{algorithmic}
\end{algorithm}

The confirmed binary mask $M^{(t)}$ is used to determine the background area at time point $t$ for loss term $\mathcal{L}_{B}$ in Eq. (\ref{eq:loss}), noting that the background area is flat and bears constant attenuation and phase shift. This provides additional leverage for the optimization problem to converge faster. This improvement is confirmed by our results in Section \ref{sec:mask} (For instance, see Fig. \ref{fig:results1} and Table \ref{tab:result}).


\subsection{Network Architecture}
The network consists of a generator $G$ and a discriminator $D$ (Fig. \ref{fig:block}). Although the proposed framework is general and any typical generative network and binary classifier can be used for $G$ and $D$; here, we provide the details of the utilized networks for the sake of completeness. We use the modified version of the auto-encoder (AE) in \cite{li2020deep} as our generator network (Fig. \ref{fig:G_SR}). 
The AE network consists of 8 convolutional layers in the encoder and 8 in the decoder part. Max pooling, and transposed convolution operators are used to perform downsampling and upsampling, respectively. One key modification we made is adding 2 more convolutional layers 1 more transposed convolutional layers to enable super-resolution, which brings further improvement at a reasonably low computation cost.

The discriminator network $D$ uses an architecture similar to the encoder part of the AE-based generator $G$. It consists of 8 convolutional layers, a global pooling layer, and a dense layer. It outputs a single value that represents the evaluation score. 
Batch Normalization \cite{ioffe2015batch} is used for both $G$ and $D$ to stabilize the training progress. The architectural details for $G$ and $D$ are given in Tables \ref{tab:str_gen} and \ref{tab:str_de}. 
To show the generalizability of the architecture, we also used DCDO \cite{niknam2021holographic}) as an alternative generator network in our experiments. 

\begin{figure}[htbp]
\begin{center}
\centerline{\includegraphics[width=0.7\columnwidth]{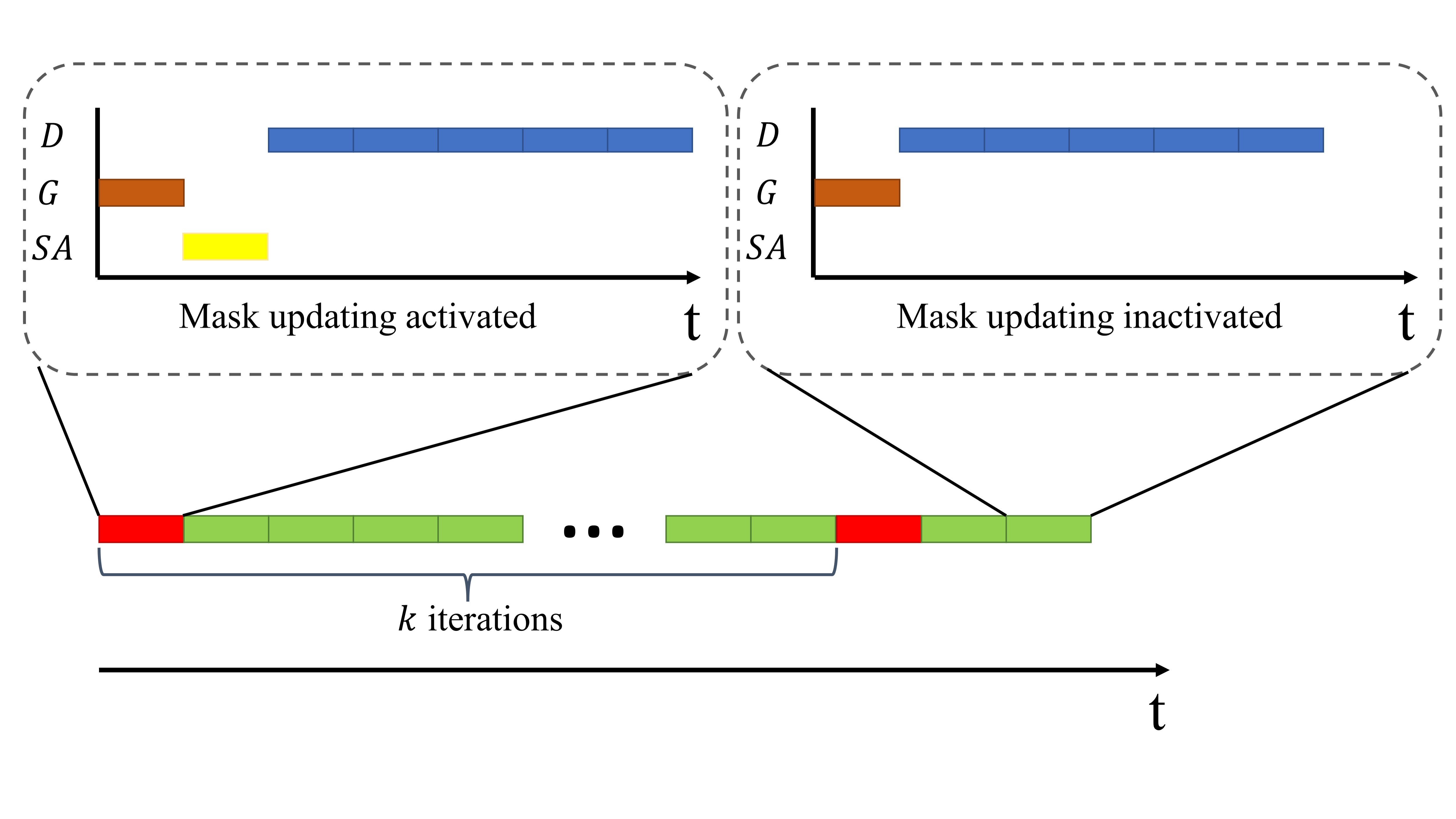}}
\caption{The training strategy. The masking update is activated once every $k=100$ intervals (shown by red). Each interval includes one iteration of the generator update (brown) followed by five iterations of the discriminator update (blue). If masking update is active (in red intervals), it is performed between the generator training and discriminator training (yellow).}
\label{fig:train}
\end{center}
\end{figure}

The training strategy is shown in Fig. \ref{fig:train}. The generator and discriminator are trained sequentially. However, to avoid the early convergence of the generator, we train the generator only once, then train the discriminator for 5 consecutive iterations. Note that the early convergence of the generator is not desirable, since any mediocre generator can produce artificial results that can fool a discriminator that has not yet reached its optimal operation. Therefore, we let the discriminator converge first and perform its best, then train the generator accordingly to produce accurate object waves from the captured holograms. 
The aforementioned masking update by the SA-based algorithm is performed after updating the generator. This does not occur after every update, but rather once after every $k$ update of the generator, as shown by the red intervals in Fig. \ref{fig:train}).

\begin{minipage}{\textwidth}

\begin{minipage}[t]{0.5\textwidth}
\makeatletter\def\@captype{table}
\resizebox{1\textwidth}{!}{%
\begin{tabular}{ll}\toprule
\textbf{Layer Type}            & \begin{tabular}[c]{@{}l@{}}\textbf{Kernel Size}\\ $K_1\times K_2\times C_{in}\times C_{out}$\end{tabular} \\ \midrule
Conv2d+BatchNorm+Relu & $5\times5\times2\times32$    \\
Conv2d+BatchNorm+Relu & $3\times3\times32\times32$   \\
MaxPool2d             &             \\
Conv2d+BatchNorm+Relu & $3\times3\times32\times64$   \\
Conv2d+BatchNorm+Relu & $3\times3\times64\times64$   \\
MaxPool2d             &             \\
Conv2d+BatchNorm+Relu & $3\times3\times64\times128$  \\
Conv2d+BatchNorm+Relu & $3\times3\times128\times128$ \\
MaxPool2d             &             \\
Conv2d+BatchNorm+Relu & $3\times3\times128\times128$ \\
Conv2d+BatchNorm+Tanh & $3\times3\times128\times16$  \\
Conv2d+BatchNorm+Relu & $3\times3\times16\times128$  \\
Conv2d+BatchNorm+Relu & $3\times3\times128\times128$ \\
ConvTranspose2d       & $stride =2 $  \\
Conv2d+BatchNorm+Relu & $3\times3\times128\times64$  \\
Conv2d+BatchNorm+Relu & $3\times3\times64\times64 $  \\
ConvTranspose2d       & $stride =2$   \\
Conv2d+BatchNorm+Relu & $3\times3\times64\times32$   \\
Conv2d+BatchNorm+Relu & $3\times3\times32\times32$   \\
ConvTranspose2d       & $stride =2$   \\
\hdashrule[0.75ex]{0.6\textwidth}{1pt}{1ex} \\
Conv2d+BatchNorm+Relu* & $3\times3\times32\times16$   \\
Conv2d+BatchNorm+Relu* & $3\times3\times16\times16$   \\
ConvTranspose2d*       & $stride =2$   \\
\hdashrule[0.75ex]{0.6\textwidth}{1pt}{1ex} \\
Conv2d+BatchNorm+Relu & $3\times3\times16\times16$   \\
Conv2d+BatchNorm+Relu & $3\times3\times16\times16 $  \\
Conv2d                & $3\times3\times16\times2$     \\ \bottomrule
\end{tabular}%
}
\caption{The architectural details of generator $G$. It utilizes a hourglass autoencoder structure. $K_1$ and $K_2$ denote the kernel size and $C_{in}$ and $C_{out}$ denote the input channel and the output channel, respectively. Layers with * are used for super-resolution.}
\label{tab:str_gen}
\end{minipage}
\begin{minipage}[t]{0.5\textwidth}
\makeatletter\def\@captype{table}
\resizebox{1\textwidth}{!}{%
\begin{tabular}{ll}\toprule
Layer Type            & \begin{tabular}[c]{@{}l@{}}\textbf{Kernel Size}\\ $K_1\times K_2\times C_{in}\times C_{out}$\end{tabular} \\ \midrule
Conv2d+BatchNorm+ReLU & $5\times5\times2\times32$    \\
Conv2d+BatchNorm+ReLU & $3\times3\times32\times32$   \\
MaxPool2d             &     -        \\
Conv2d+BatchNorm+ReLU & $3\times3\times32\times64$   \\
Conv2d+BatchNorm+ReLU & $3\times3\times64\times64$   \\
MaxPool2d             &     -        \\
Conv2d+BatchNorm+ReLU & $3\times3\times64\times128$  \\
Conv2d+BatchNorm+ReLU & $3\times3\times128\times128$ \\
MaxPool2d             &     -        \\
Conv2d+BatchNorm+ReLU & $3\times3\times128\times128$ \\
Conv2d+BatchNorm      & $3\times3\times128\times16$  \\
GlobalPool2d          &     -        \\
Full-connected        & $1\times1\times16\times1$   \\ \bottomrule
\end{tabular}%
}
\caption{The architectural details of discriminator $D$. It outputs the similarity of the input and the target. }
\label{tab:str_de}
\end{minipage}
\end{minipage}

%% file: experiment.tex
\section{Experiment}
In this section, we verify the performance of the proposed algorithm using simulated holograms, publicly available samples, and our \textit{dendritic} tags.

\subsection{Experiment Setup}





\begin{figure}[htbp]
\centering\includegraphics[width=1\linewidth]{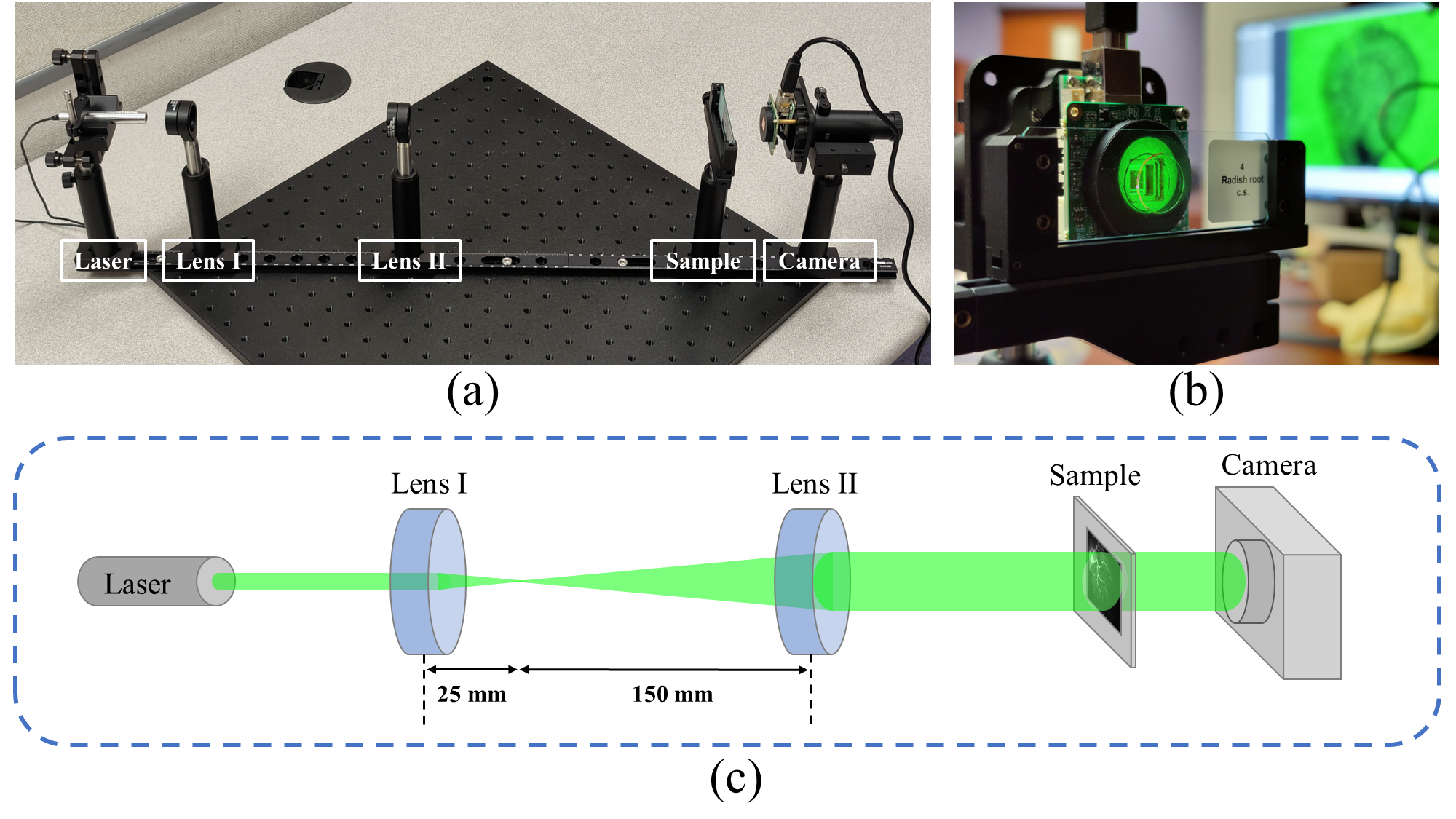}
\caption{(a) Utilized experimental setup for in-line holography, (b) sample test, (c) using two lens to enlarge the beam intersection.}
\label{fig:lab_1}
\end{figure}

Our experiment setup is shown in Fig. \ref{fig:lab_1}. The laser module CPS532-C2 is used to generate a single wavelength (532 $\bf{nm}$) laser beam with a round shape of diameter 
3.5 $\bf{mm}$. The laser module provides 0.9 $\bf{mW}$ a typical USB port power. The USB-based powering facilitates taking clear holograms in normal conditions. 
We use a digital camera A55050U, which employs a $1/2.8^"$ \textit{Complementary Metal-Oxide Semiconductor} (CMOS) sensor with 2.0 $\bf{\mu m}$ $\times$ 2.0 $\bf{\mu m}$ pixel size. This sensor provides picture quality with 22 frames per second (fps) at a resolution of 5 Mega-Bytes (2560 $\times$ 1920 pixels), which gives a 5120 $\bf{\mu m}$ $\times$ 3840 $\bf{\mu m}$ field Of view (FOV). Rolling shutter and variable exposure time also provide convenience for fast and accurate imaging. Note that this architecture can be made compact by substantially lowering the distances for portable readers.

As shown in Fig. \ref{fig:lab_1}(c), two convex lenses with focal lengths of $f_1=$ 25 $\bf{mm}$ and $f_2=$ 150 $\bf{mm}$ are applied to expand the laser beam, so that the laser beam fully covers the dendrite samples. The lenses located at distance $f_1+f_2=175 \bf{mm}$ from one another, so their focal points collocate to retain the plane wavefront. The magnifying power of this system is $MP = \frac{f2}{f1} = \frac{150}{25} = \mathbf{6}$, which enlarges the laser intersection diameter from 3.5 $\bf{mm}$ to 21 $\bf{mm}$.
In Fig. \ref{fig:lab_1}(b), a sample slide is placed on the sample holder; the laser beam passes through the sample, and propagates the hologram onto the sensor plane. The captured image is displayed on the computer in real-time and is fed to the proposed DL-based recovery algorithm. 
With an exposure time of 28 $\mu \bf{s}$, the hologram is captured in clear and bright conditions. 

The DL framework is developed in Python environment using the Pytorch package and Adam optimizer. Training is performed using two Windows 10 machines with NVIDIA RTX2070 and RTX3090 graphics cards.

\subsection{Dendrite Samples}

\begin{figure}[h]
\begin{center}
\centerline{\includegraphics[width=0.6\columnwidth]{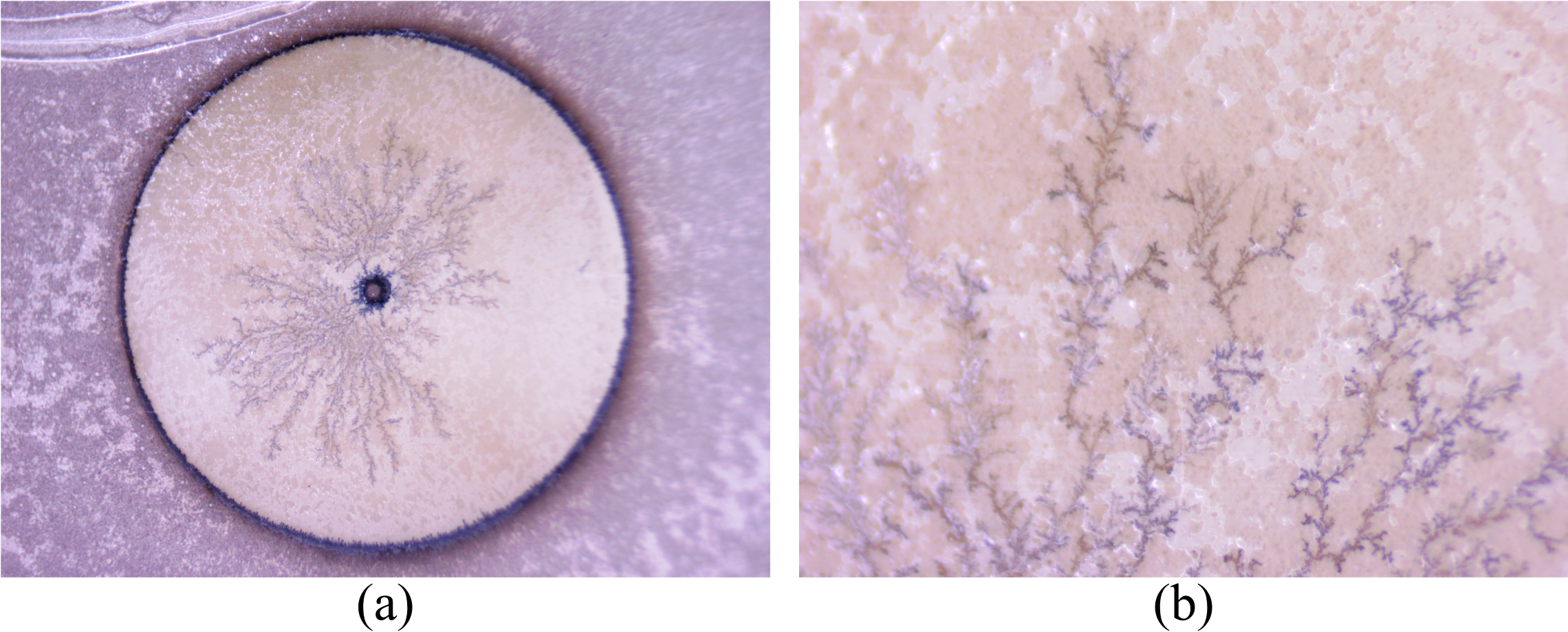}}
\caption{A \textit{dendritic} pattern grown on a synthetic paper and 
soaked with a liquid electrolyte.}
\label{fig:dendrite}
\end{center}
\end{figure}

In addition to simulated and public holograms, we use \textit{dendrite} samples in our experiments. \textit{Dendrites} are visual identifiers that are formed by growing tree-shaped metallic fractal patterns by inducing regulated voltage on electrolyte solutions with different propensities \cite{kozicki2021dendritic}. These tags can be efficiently produced in large volumes on multiple substrate materials (e.g., mica, synthetic paper, etc.) with different granularity and density \cite{kozicki2022dendritic}. 
A \textit{dendrite} sample is shown in Fig. \ref{fig:dendrite}.
\textit{Dendrites} have specific features such as extremely high entropy for their inherent randomness, self-similarity, and unclonability due to their 3D facets and non-resolution granularity.
These features make this patented technology an appropriate choice for security solutions, including identification tags, visual authentication, random generators, and producing physical unclonable functions (PUFs) with robust security keys \cite{razi2022methods,chi2020consistency}.

We previously have shown \textit{dendrites}' utility as 2D authentication identifiers \cite{valehi2017graph,chi2020consistency}, but exploiting information-rich features from \textit{dendrites} to achieve unclonability requires specific technologies such as digital holography, as presented in our previous work \cite{li2020deep}.


\subsection{Test with Simulated Holograms}\label{sec:simu}

First, we compare the performance of our method against the most powerful untrained methods, where the sample hologram is sufficient to recover the phase with no need for a training dataset. 
It is noteworthy that in general, there exist two main classes of untrained neural networks, one with encoder-decoder architecture, mainly based on \textit{deep autoencoders}, (e.g., DeepDIH \cite{li2020deep}), 
and another class with only the decoder part, 
the so-called \textit{deep decoder} (e.g., DCDO \cite{niknam2021holographic}). 

In this experiment, we compare our model with two untrained DL methods (DeepDIH and DCDO) as well as a CS-based method proposed in \cite{zhang2018twin} using USAF target samples. In our framework, we use the fine-tuned version of DeepDIH as the generator network, but we also perform ablation analysis by replacing it with the DCDO.

\begin{figure}[htbp]
\begin{center}
\centerline{\includegraphics[width=0.8\columnwidth]{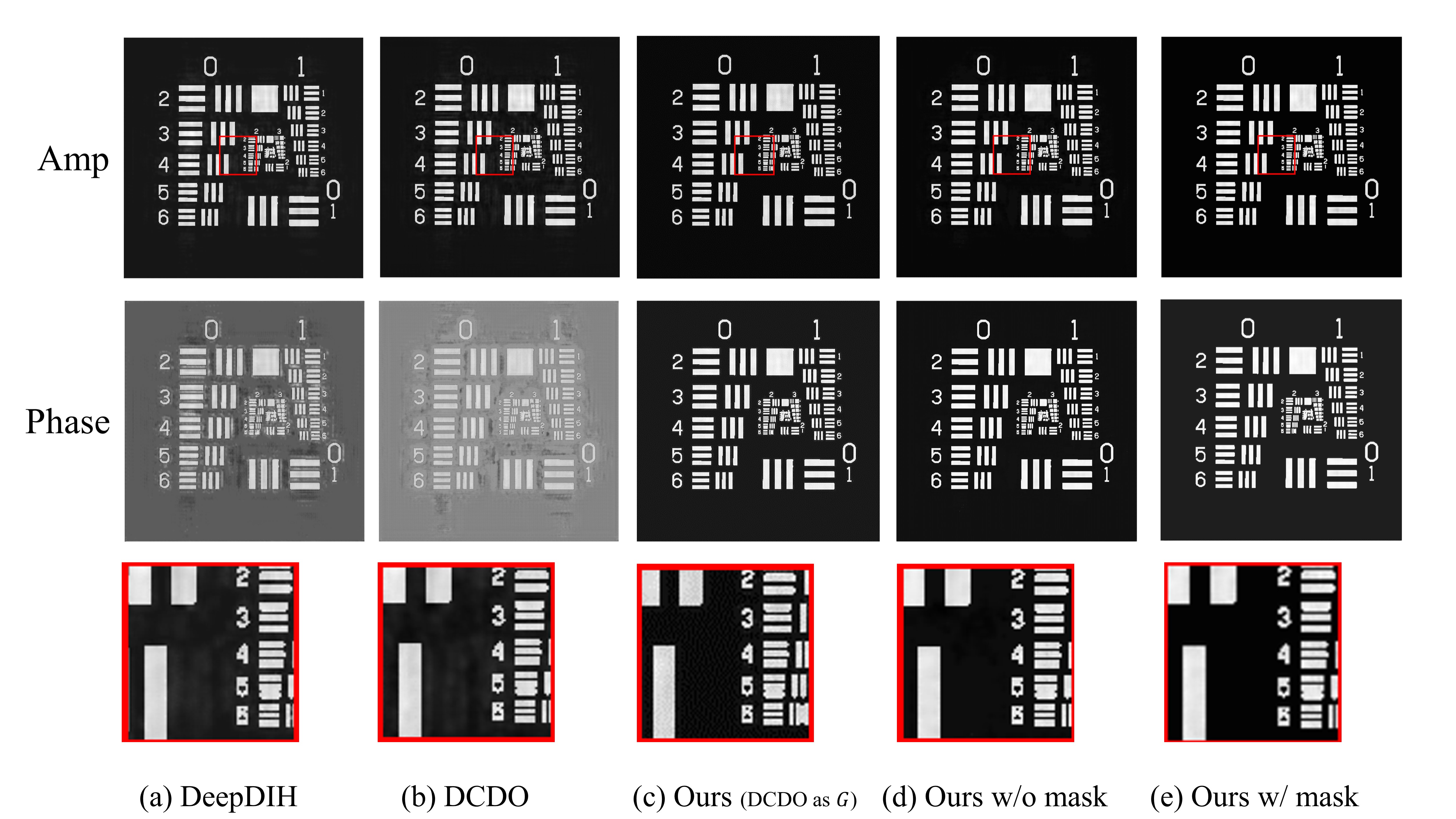}}
\caption{The comparison of different methods, including 
(a) DeepDIH \cite{li2020deep}, (b) DCDO \cite{niknam2021holographic}, (c) proposed method using DCDO as generator, (d) proposed method with modified DeepDIH as generator, and (e) same as (d) with adaptive masking module. First, second, and third rows represent the reconstructed amplitude, phase, and amplitude of select zone, respectively.}
\label{fig:results1}
\end{center}
\end{figure}

The results in Fig. \ref{fig:results1} and Table \ref{tab:result} demonstrate the superiority of our proposed method. Particularly, the PNSR of our method ranges from 25.7 dB to 29 dB, depending on the choice of the generator and activating/inactivating the adaptive masking module, which is significantly higher than the CS method (PSNR 14.6 dB), DeepDIH (PNSR 19.7 dB), and DCDO (PNSR 20.1 dB). A similar observation is made in Fig. \ref{fig:results1}, especially in the quality of the reconstructed object phase.   
The main justification for this huge improvement is that 
the untrained method with deep autoencoder without proper regularization terms can easily be trapped in overfitting the noise, especially if over-parameterized \cite{heckel2018deep}. Although the DCDO method uses fewer parameters to alleviate the overfitting issue, it does not employ complete knowledge about the hologram formation process and uses random input. 
In contrast, our method uses the back-propagated holograms as the generator input, meaning that the generator network training starts from a reasonably good start point and converges to a better optimum.

Another drawback of the competitor methods is using MSE loss which does not adequately capture the image reconstruction quality and may guide the network to converge wrongly. This issue is solved in our method by leveraging the underlying physics law and using a learnable distance measure through the discriminator network. 

Finally, we observe a significant improvement for the utilized adaptive masking module that improves the reconstruction quality from PSNR 26.3 dB to as high as 29 dB. This highlights the advantage of incorporating physical knowledge into the reconstruction process by adding more constraints to the network weights through background loss. 

\begin{figure}[htbp]
\begin{center}
\centerline{\includegraphics[width=0.8\columnwidth]{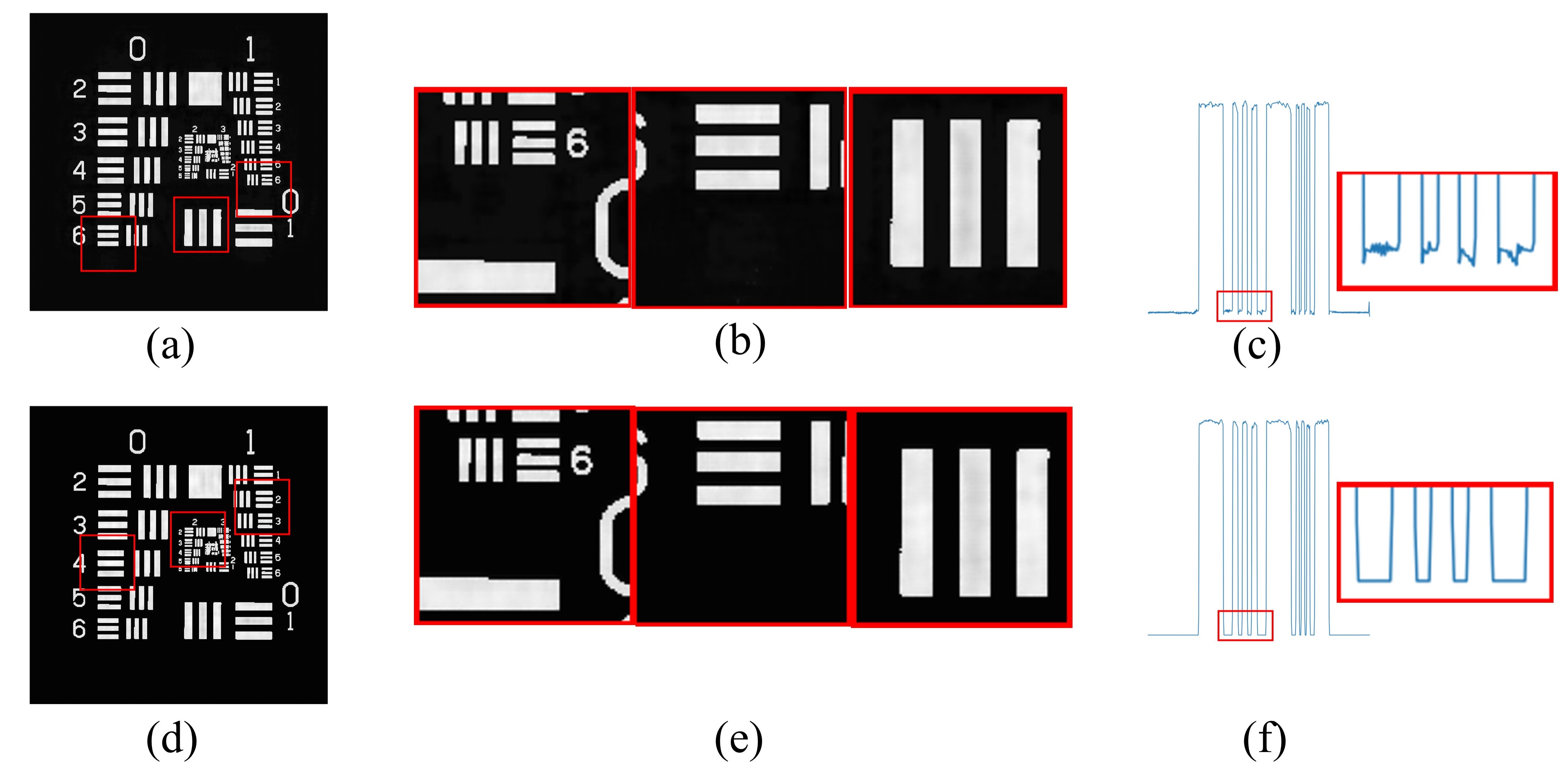}}
\caption{The comparison of the reconstructed object wave from captured hologram using the proposed model without imposing background loss (top row) and with background loss (bottom row). 
Left (a,d): amplitude; Middle (b,e): zoom-in details of amplitude; Right (c,f): side view of one row of the object blades' surface.} 
\label{fig:maskvs}
\end{center}
\end{figure}

Fig. \ref{fig:maskvs} provides a closer look at the benefits of using the adaptive masking module and applying background loss to USAF target. For a better visibility, we compare three selected parts of the reconstructed amplitude (middle) and the side-view of the reconstructed object surface. It is clearly seen that imposing background loss smooths out the background part of the image and improves the reconstruction quality while not causing edge distortion damage.

\begin{table}[]
\centering
\resizebox{0.8\textwidth}{!}{%
\begin{tabular}{ccccccc}\toprule
\textbf{Method} & CS      & DIH         & DCDO      & \textbf{Ours} (DCDO as G) & \textbf{Ours} w/o mask & \textbf{Ours} w/ mask \\ \midrule
\textbf{PSNR}(dB)   & 14.590 & 19.657 & 20.056 & 25.728             & 26.325         & 29.019 \\ \bottomrule
\end{tabular}%
}
\caption{The comparison of different methods, including compressive sensing (CS) method \cite{zhang2018twin}, DeepDIH \cite{li2020deep}, DCDO \cite{niknam2021holographic}, proposed method with DCDO as generator, and proposed method with modified DeepDIH as generator without and with adaptive masking module.}
\label{tab:result}
\end{table}

\begin{figure}[htbp]
\begin{center}
\centerline{\includegraphics[width=1\columnwidth]{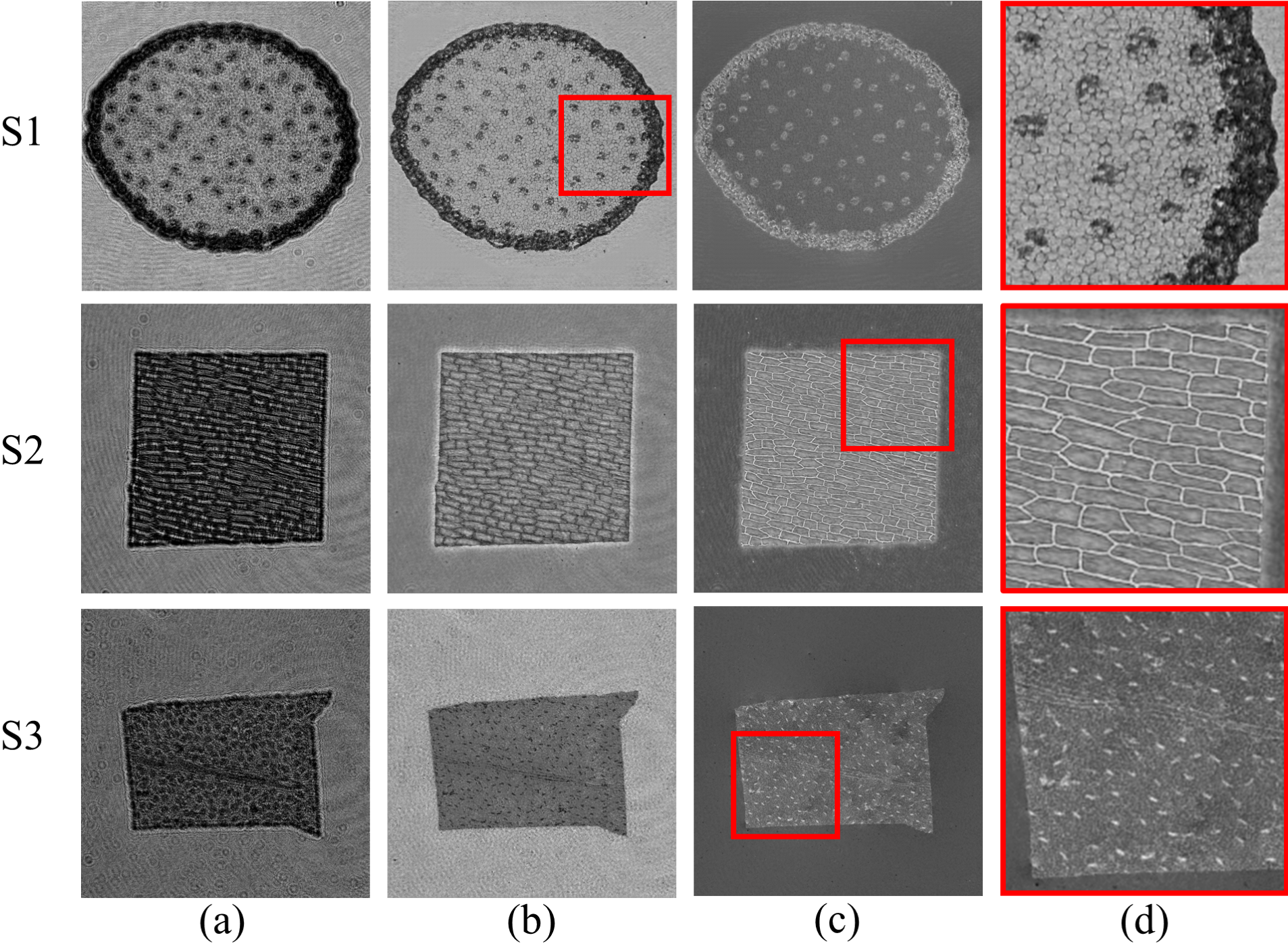}}
\caption{The reconstruction of the three real samples S1: Zea Stem, S2: Onion Epidermis, S3: Stomata-Vicia, Faba Leaf). (a) captured hologram; (b) reconstructed amplitude; (c) reconstructed phase; (d) zoom in part.}
\label{fig:cell_1}
\end{center}
\end{figure}

\subsection{Test with Real Samples}\label{sec:dendrite}

To prove the applicability of our model in real-world scenarios, 
we have tested different types of samples, including S1: Zea Stem, S2: Onion Epidermis, and S3: Stomata-Vicia Faba Leaf (Fig. \ref{fig:cell_1}).
The average cell sample size is 2 $\bf{mm}$ $\times$ 2 $\bf{mm}$, equivalent to 1000 $\times$ 1000 pixels in the sensor field. 
All samples have been placed at a distance of 5.5 $\bf{mm}$ (the closest possible) to the CMOS sensor to avoid unnecessary diffraction of the object waves \cite{niknam2021holographic}.
The parameters of the framework are set accordingly. For example, we set pixel size (2 $\bf{\mu m}$), wavelength (0.532 $\bf{nm}$), and the distance from sample to sensor (5,500 $\bf{\mu m}$).

\begin{figure}[htbp]
\begin{center}
\centerline{\includegraphics[width=1\columnwidth]{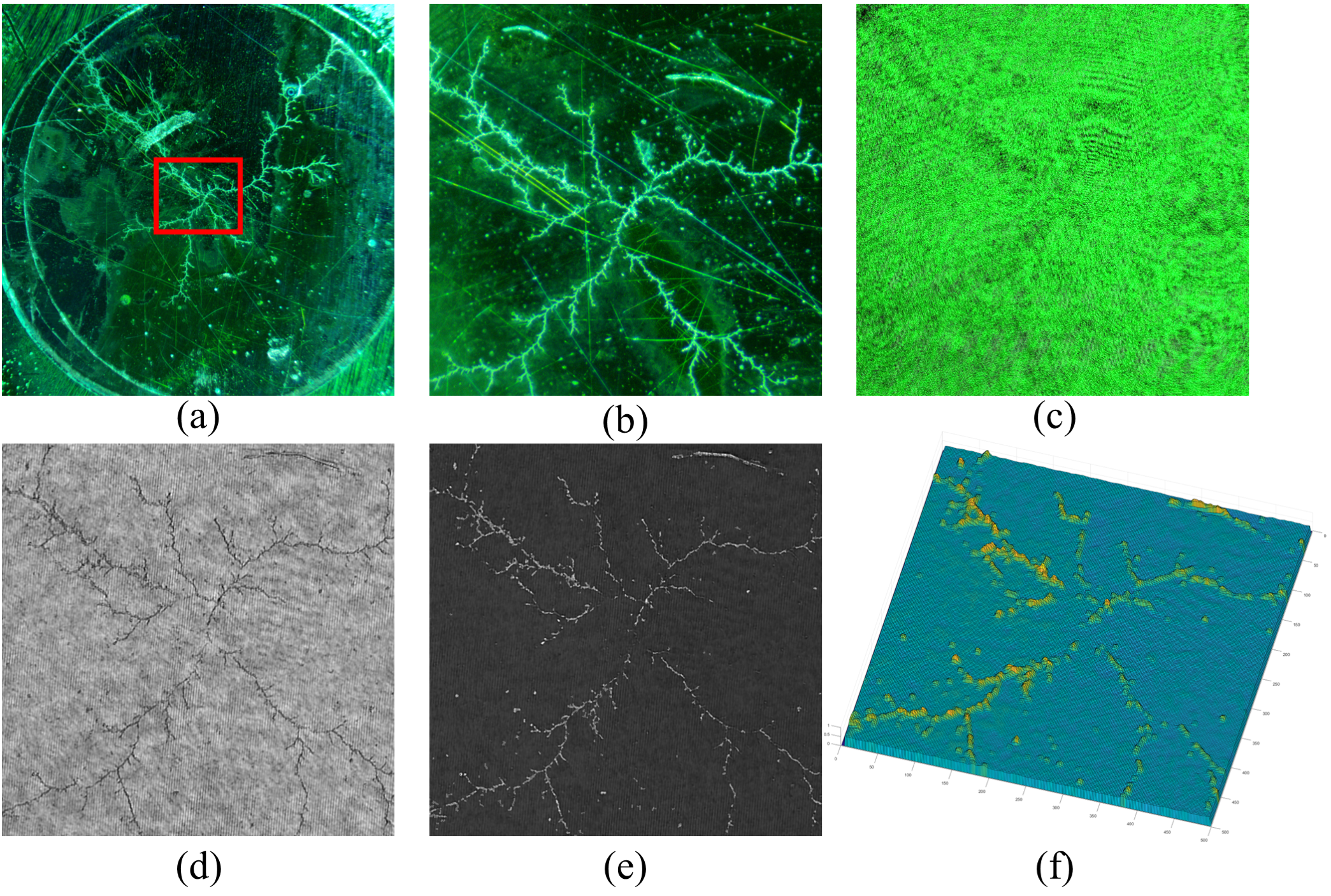}}
\caption{The reconstruction process of a dendrite sample. (a) a typical mica-substrate dendrite sample; (b) select part of the dendrite; (c) captured hologram; (d) reconstructed amplitude; (e) reconstructed phase; (f) 3D view of the reconstructed object surface.}
\label{fig:dendrite_holo}
\end{center}
\end{figure}

We also used the same setup to capture holographic readings of \textit{dendrite} samples (Fig. \ref{fig:dendrite_holo}). 
The results are presented after convergence which occurs after 2,000 epochs. 
The results in Figs. \ref{fig:cell_1} and \ref{fig:dendrite_holo} demonstrate the end-to-end performance of the proposed GAN-based phase recovery when applied to real holograms captured by our DIH setup.

\subsection{Robustness to Noise}
\begin{figure}[htbp]
\begin{center}
\centerline{\includegraphics[width=0.8\columnwidth]{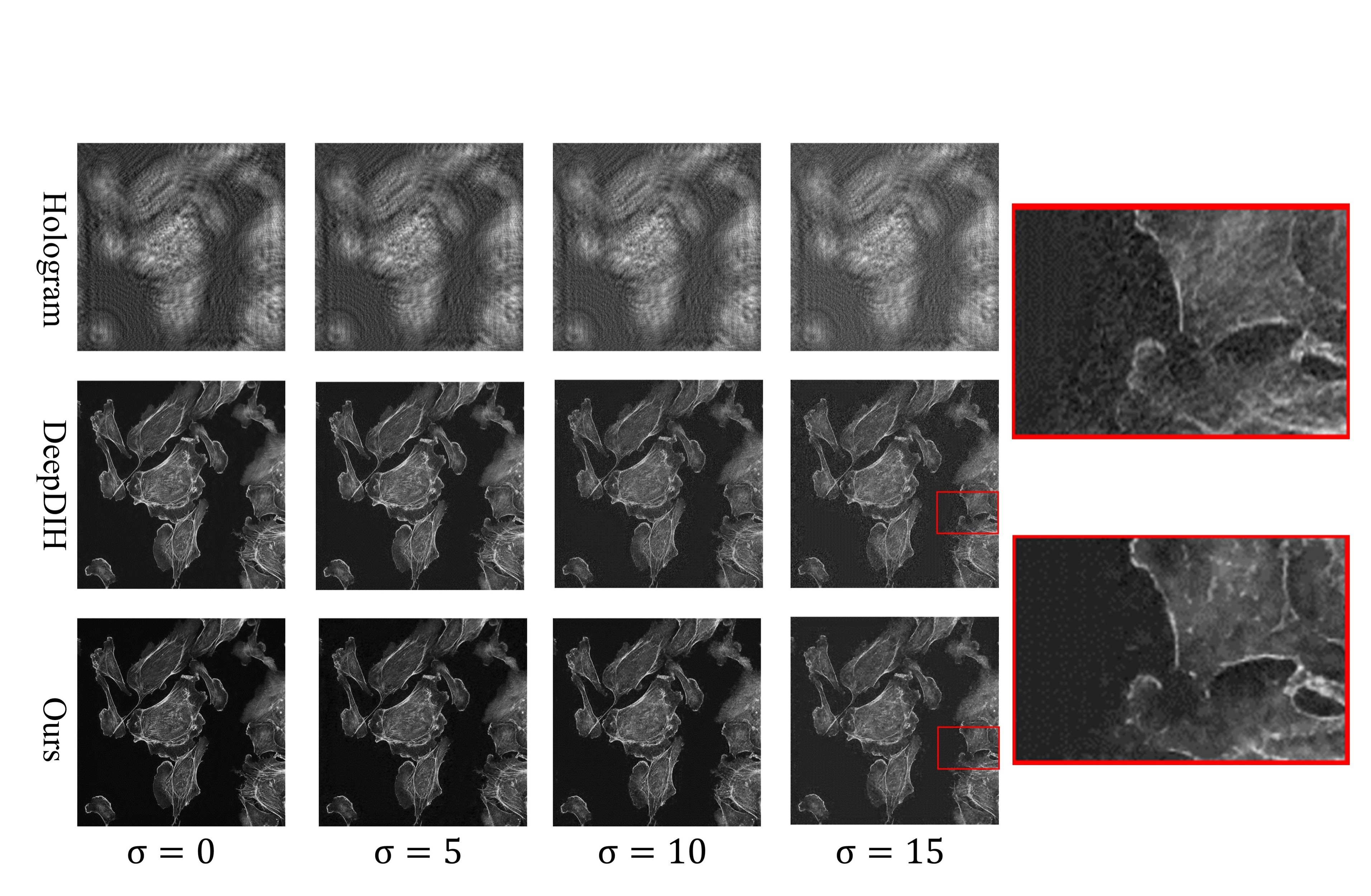}}
\caption{Reconstructed amplitude of the cell sample. The first row is the simulated hologram under noise level 0, 5, 10, and 15, respectively. } 
\label{fig:noise1}
\end{center}
\end{figure}

\begin{figure}[htbp]
\begin{center}
\centerline{\includegraphics[width=0.8\columnwidth]{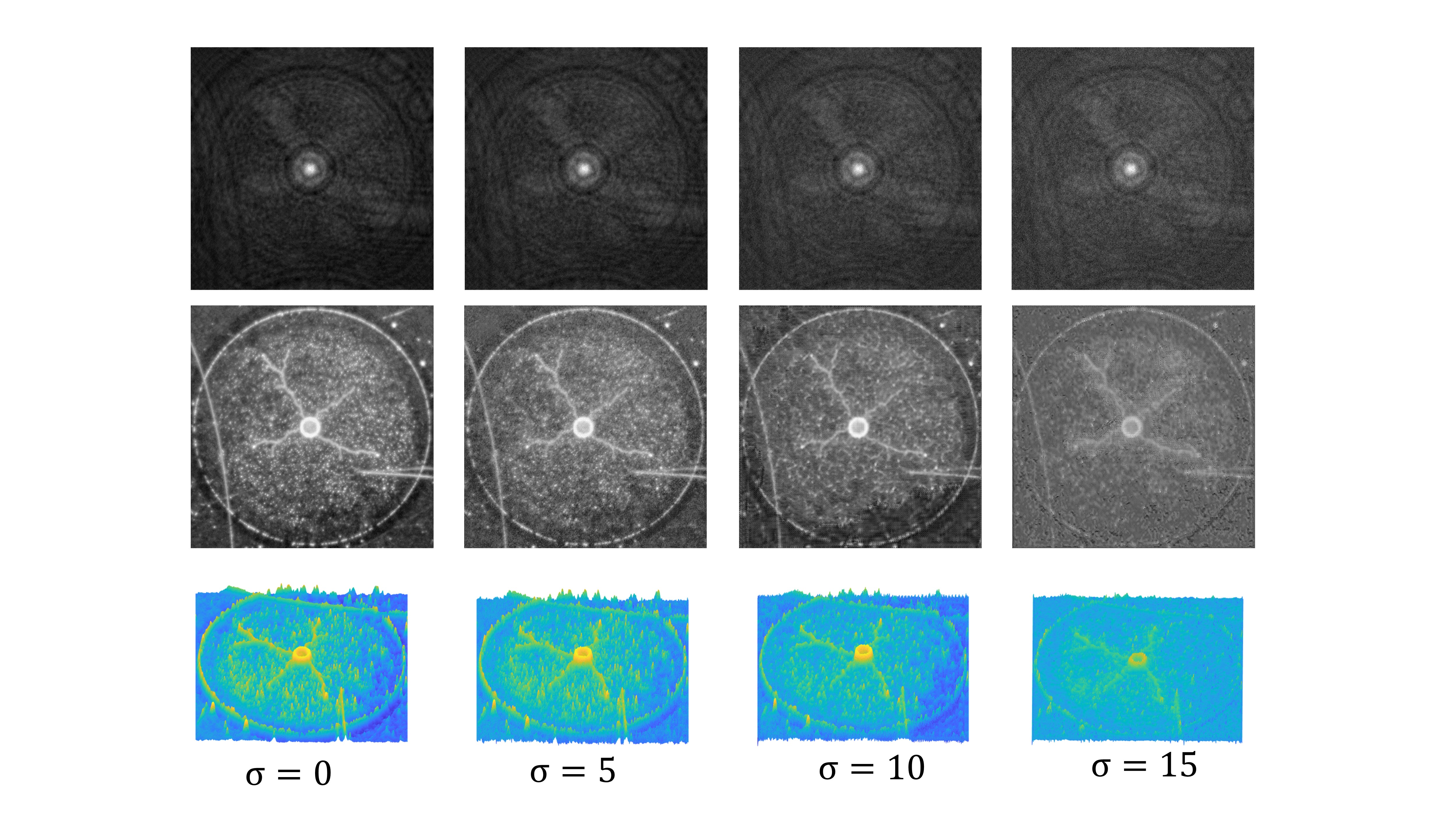}}
\caption{Reconstructed phase of the dendrite sample with their 3D plot. The first row is the captured hologram with artificially added noise with standard deviations $\sigma=0, \sigma=5, \sigma=10$, and $\sigma=15$, respectively.} 
\label{fig:noise2}
\end{center}
\end{figure}

Like regular images, the holographic readings can be noisy due to the illumination conditions, rusty lens, sensor noise, and other imaging artifacts ($\epsilon$ in Eq. (\ref{eq:2})). 
We examine noise impact to ensure reasonable noise levels do not substantially degrade the reconstruction quality. To this end, we intentionally add additive white Gaussian noise (AWGN) of different levels (standard deviation: $\sigma=5$, $\sigma=10$, and $\sigma=5$) to the captured holograms. The results for cell and dendrite samples are respectively presented in Figs. \ref{fig:noise1} and \ref{fig:noise2}, and summarized in Table \ref{tab:noise}).

The results in Figs. \ref{fig:noise1} 
and \ref{fig:noise2}
show that the phase recovery of our algorithm is fairly robust against noise levels up to $\sigma=10\sim 15$, and significantly improves upon the DeepDIH framework. 
According to Table \ref{tab:noise}, by increasing the noise up to $\sigma=10$, the performance decay of our method is smaller than that of the DeepDIH method.
For instance, for the cell sample, DeepDIH shows around 3 dB decay for each $\delta\sigma=5$ increase in the noise level, while ours only shows around 2 dB decay. This represents a 50\% improvement in PSNR vs noise increase rate.
In the dendrite sample, from $\sigma=5$ to $\sigma=10$, the SSIM of DeepDIH decreases about 0.2, while that of ours only decreases about 0.06, which is $70\%$ smaller. We declare conservatively that the reconstruction quality is acceptable for noise levels up to $\sigma=10$, which incurs only around 4 dB decay in PSNR and around 0.2 SSIM loss.

The results overall confirm the robustness of the proposed model for noisy images. Part of this robustness is inherited from the intrinsic noise removal capability of AEs used as a generator in our framework. Also, imposing TV loss on the background section of the hologram removes high-frequency noise from the image.

\begin{table}[htbp]
\centering
\resizebox{0.5\textwidth}{!}{%
\begin{tabular}{ccccccc}\toprule
\multicolumn{1}{l}{}           & \multicolumn{2}{c}{\textbf{Noise Level ($\sigma$)}}          & 0      & 5      & 10     & 15     \\ \midrule
\multirow{4}{*}{\textbf{Cell}} & \multirow{2}{*}{\textbf{DeepDIH}} & \textbf{PSNR} & 26.008 & 22.933 & 20.663 & 18.311 \\
                               &                                   & \textbf{SSIM} & 0.807  & 0.699  & 0.585  & 0.491  \\
                               & \multirow{2}{*}{\textbf{Ours}}    & \textbf{PSNR} & 27.817 & 26.062 & 23.732 & 19.793 \\
                               &                                   & \textbf{SSIM} & 0.941   & 0.8418 & 0.716  & 0.548  \\ \midrule
\multirow{4}{*}{\textbf{Dendrite}}      & \multirow{2}{*}{\textbf{DeepDIH}} & \textbf{PSNR} & 30.113 & 29.207  & 21.784 & 16.720  \\
                               &                                   & \textbf{SSIM} & 0.916  & 0.875    & 0.671   & 0.453  \\
                               & \multirow{2}{*}{\textbf{Ours}}    & \textbf{PSNR} & 32.994 & 30.071 & 28.092 & 23.976 \\
                               &                                   & \textbf{SSIM} & 0.969  & 0.911  & 0.846  & 0.763 \\ \bottomrule
\end{tabular}%
}
\caption{Comparison of DeepDIH and proposed method in reconstructing phase under different noise levels.}
\label{tab:noise}
\end{table}

\subsection{One-shot Training and Transfer Learning}\label{sec:transfer}

A key challenge of DL-based phase recovery methods compared to conventional numerical methods is their generalizability and transferability to other experiments due to DL methods' unexplainability and black-box nature. This matter can be problematic in real-time applications since the time-consuming training phase should be repeated for every new sample. 
The proposed method partially alleviates this issue due to incorporating the underlying physics laws.

To investigate the transferability of our method, we develop an experiment with the following three testing scenarios for simulated holograms for 4 randomly selected \textit{neuro samples} taken from the CCBD dataset \cite{P1170}. 

\begin{enumerate}[I]
    \item "One-shot Train" model: The hologram of sample S1 is used to train the DH-GAN model and reconstruct S1 amplitude and phase as usual with 3,000 iterations. This generator part of the model is used to reconstruct the amplitude and phase of the holograms of samples S2-S4 (one model for all).  
    
    \item "Retrain:500" model: the network is initialized with random weights, then the reconstruction is performed independently for each sample using 500 iterations (four different models, one for each sample). 
    
    \item "One-shot Train+500" model: we use the model trained for sample S1 to initialize the network for other samples S2-S4, then perform reconstruction with extra 500 iterations for each sample separately. 
\end{enumerate}

\begin{figure}[htbp]
\begin{center}
\centerline{\includegraphics[width=1\columnwidth]{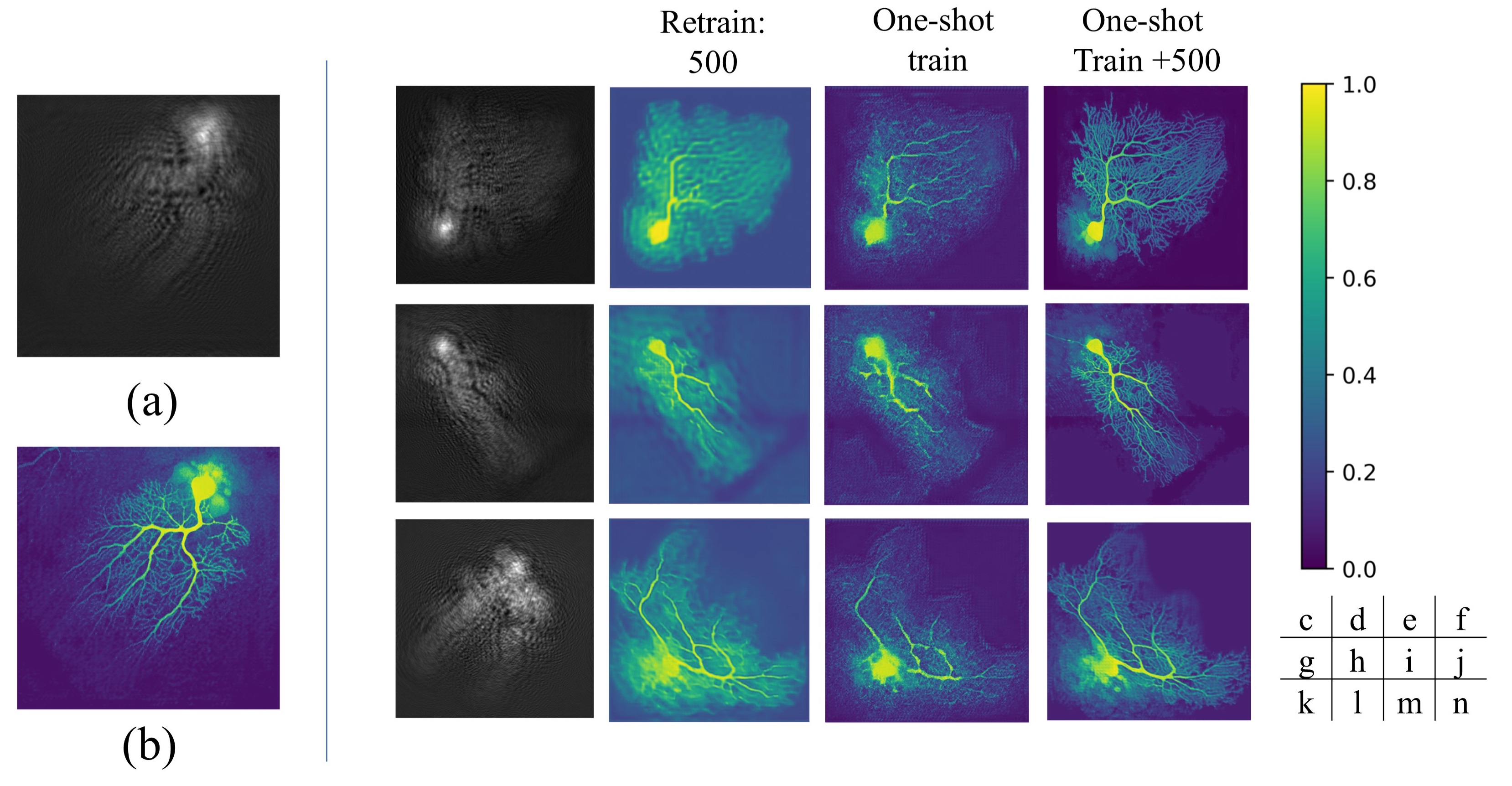}}
\caption{The transferability of the trained model. (a) captured hologram of sample S1; (b) reconstructed phase of sample S1 using the trained model. Left side: each row represents a sample (S2,S3,S4); first column represents the captured hologram, and the next three columns represent the results of the three testing scenarios.}
\label{fig:oneshot}
\end{center}
\end{figure}

The results of this experiment are shown in Fig. \ref{fig:oneshot} and Table \ref{tab:transfer_learning}.
The results in Fig. \ref{fig:oneshot}(a,b) show the excellent reconstruction quality of DH-GAN with 3,000 iterations. However, for fast deployment, one may not afford to repeat 3,000 training iterations for every new sample. In this case, one potential solution would be using the trained network for other samples (e,i,m). 
The results are quite impressive (PSNR is in 19 dB to 20.2 dB range) and can be acceptable for many applications. Indeed, it outperforms independent networks trained for new samples using only 500 training iterations and random initialization (d,h,i), which achieve a PSNR in 12.8 dB to 14.3 dB range. One intermediate solution would be transfer learning, namely using the network trained for S1 as initialization for other networks and performing 500 training iterations for new samples (f,j,n), which offers the best results (PSNR in 25 dB to 30 dB range).



\begin{table}[htbp]
\centering
\resizebox{0.5\textwidth}{!}{%
\begin{tabular}{lcccc}\toprule
      \textbf{Samples}             & \multicolumn{1}{c}{\textbf{S1}} & \multicolumn{1}{c}{\textbf{S2}} & \multicolumn{1}{c}{\textbf{S3}} & \multicolumn{1}{c}{\textbf{S4}} \\ \midrule
Retrain: 500       &       -                & 13.028                & 14.240              & 12.866                \\
One-shot Train     & 31.286           & 19.772             & 20.261             & 18.998                \\
One-shot Train+500 &      -                 & 29.476             & 25.162              & 25.576          \\ \bottomrule  
\end{tabular}%
}
\caption{The performance (in PSNR) of three transfer learning scenarios, presented in Fig. \ref{fig:oneshot}.} 
\label{tab:transfer_learning}
\end{table}

\begin{figure}[htbp]
\begin{center}
\centerline{\includegraphics[width=0.5\columnwidth]{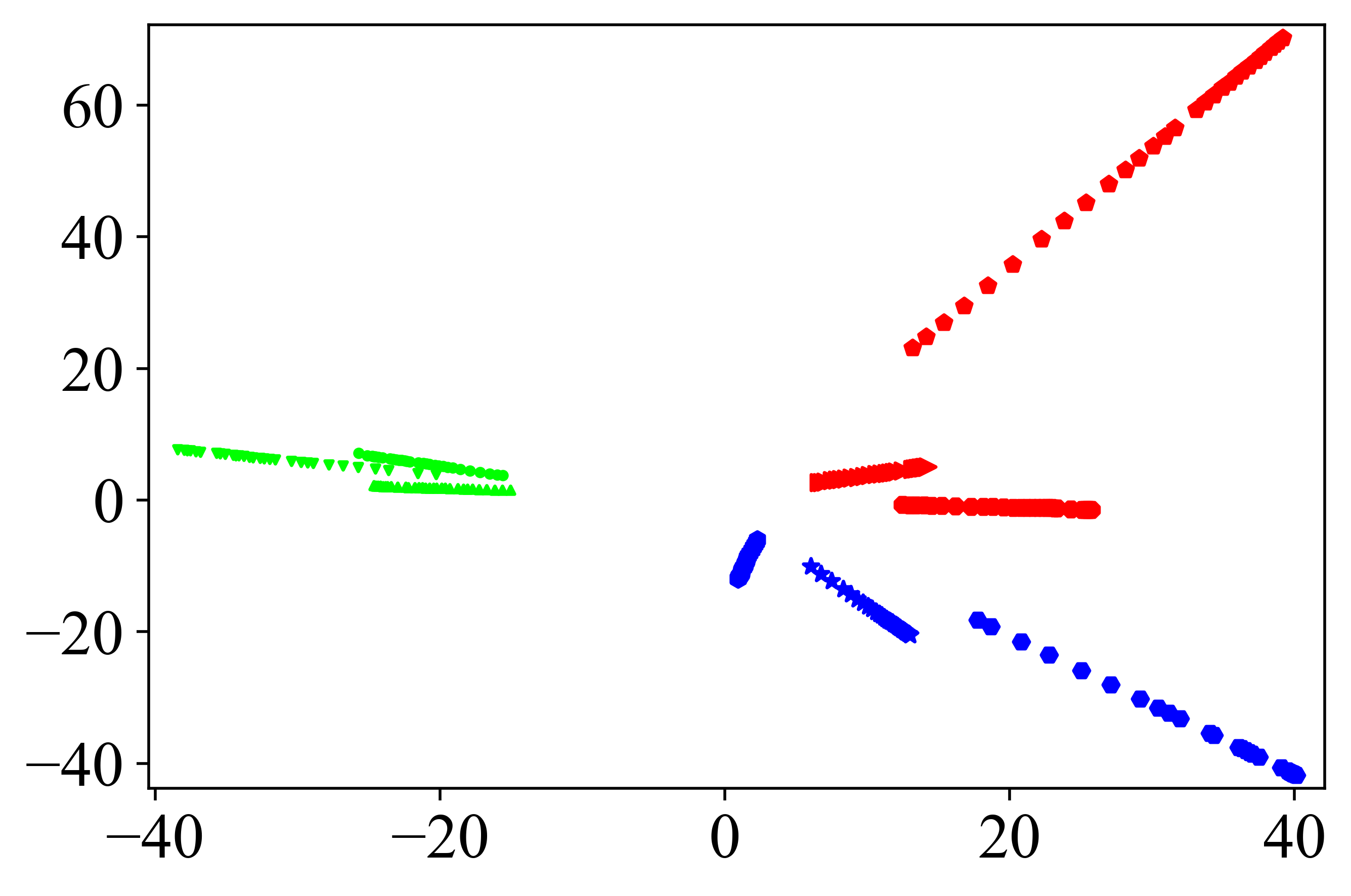}}
\caption{The 2D visualization of the network weights using PCA. Each data point represent the vector of network weights after 100 iterations. Different colors represent different sample types, including MNIST handwriting digits (green), CCBD (blue), and USAF Target (red). The weights trained for similar patterns are radially clustering with the same orientation.}
\label{fig:weights}
\end{center}
\end{figure}

To further investigate the transferability of the developed framework, we perform a test using three sample types, including:1) MNIST handwriting digits, 2) CCBD, and 3) USAF Target. We choose four samples of each type, and train an independent network with fixed initialization for each sample using 3,000 iterations. 
The weights are collected once per 100 iterations and considered a data point. Fig. \ref{fig:weights} visualizes the resulting network weights in the 2D domain using principal component analysis (PCA). The observation is quite interesting since the network weights corresponding to the sample type are aligned in the same direction, and different sample types are well separated into disjoint clusters. The practical implication of these results is that the network trained for one specific sample type can be used for similar samples of the same type but not for other sample types.

